\pdfoutput=1
\RequirePackage{ifpdf}
\ifpdf 
\documentclass[pdftex]{sigma}
\else
\documentclass{sigma}
\fi

\numberwithin{equation}{section}

\newtheorem{Theorem}{Theorem}[section]
\newtheorem{Proposition}[Theorem]{Proposition}

\newcommand{\one}{\mbox{$1 \hspace{-1.0mm} {\bf l}$}}

\begin{document}


\renewcommand{\thefootnote}{$\star$}

\newcommand{\arXivNumber}{1603.01902}

\renewcommand{\PaperNumber}{070}

\FirstPageHeading

\ShortArticleName{Flowing in Group Field Theory Space: a Review}

\ArticleName{Flowing in Group Field Theory Space: a Review\footnote{This paper is a~contribution to the Special Issue on Tensor Models, Formalism and Applications. The full collection is available at \href{http://www.emis.de/journals/SIGMA/Tensor_Models.html}{http://www.emis.de/journals/SIGMA/Tensor\_{}Models.html}}}

\Author{Sylvain CARROZZA}

\AuthorNameForHeading{S.~Carrozza}

\Address{Universit\'e Bordeaux, LaBRI, UMR 5800, 33400 Talence, France}
\Email{\href{mailto:sylvain.carrozza@labri.fr}{sylvain.carrozza@labri.fr}}

\ArticleDates{Received March 08, 2016, in f\/inal form July 13, 2016; Published online July 16, 2016}

\vspace{-1.5mm}

\Abstract{We provide a non-technical overview of recent extensions of renormalization methods and techniques to Group Field Theories (GFTs), a class of combinatorially non-local quantum f\/ield theories which generalize matrix models to dimension $d \geq 3$. More precisely, we focus on GFTs with so-called closure constraint, which are closely related to lattice gauge theories and quantum gravity spin foam models. With the help of recent tensor model tools, a rich landscape of renormalizable theories has been unravelled. We review our current understanding of their renormalization group f\/lows, at both perturbative and non-perturbative levels.}

\Keywords{group f\/ield theory; quantum gravity; quantum f\/ield theory; renormalization}

\Classification{81T15; 81T16; 83D27; 83C45}

\renewcommand{\thefootnote}{\arabic{footnote}}
\setcounter{footnote}{0}

\vspace{-3.5mm}

\section{Introduction}

\looseness=-1
From a mathematical perspective, a GFT \cite{ad_ten, freidel_gft, Krajewski_rev, daniele_rev2006, daniele_rev2011} is a quantum f\/ield theory def\/ined on $d$ copies of a compact Lie group $G$, in which point-like interactions are replaced by non-trivial combinatorial objects. At the level of the f\/ield theory action, this translates into peculiar non-localities of the interactions, which are given by pairwise identif\/ications and integrations of individual group variables of the elementary f\/ields. One could thus say that a given GFT interaction is local in each pair of copies of the group~$G$ thus identif\/ied, but non-local from the point of view of the full conf\/iguration space~$G^d$. This leads to subtleties in the construction and analysis of such models, which for a~long time precluded the def\/inition of renormalizable theories.

Alternatively, GFTs can be seen as generalized matrix and tensor models, with group-valued rather than discrete indices. Progress in tensor models (see, e.g., \cite{uncoloring, double_scaling, universality, razvan_jimmy_rev} and references therein) can therefore be directly imported into the GFT formalism. The continuous structure of the group allows to considerably enrich this purely combinatorial background and def\/ine more general classes of tensorial theories, which go under the name of Tensorial Group Field Theories (TGFTs) or simply Tensorial Field Theories (TFTs). The introduction of derivative couplings, and in particular of non-trivial propagators, leads to a f\/irst class of such models
\cite{josephaf,joseph_d2,Geloun:2015lta,joseph_etera, riccardo, tensor_4d, josephsamary, joseph_reiko, samary_2point, dine_fabien_2015}. They are proper f\/ield theories whose (perturbative) def\/initions already require a non-trivial renormalizability analysis. Since the group plays a~limited role and enjoys no particular physical interpretation in such models, they are preferably referred to as TFTs. Second, the group structure can be used to impose particular constraints on the elementary f\/ields, which endow Feynman amplitudes with the structure of generalized lattice gauge theories. This latter class of models is usually referred to as Tensorial Group Field Theories (TGFTs) to emphasize the central importance of the group\footnote{Note however that the distinction between TFTs and TGFTs is somehow lose, as it relies more on questions of intent and interpretation than on a mathematically precise def\/inition. These terms are therefore sometimes used interchangeably in the literature. In the present review, we will also refer to TFTs as \emph{TGFTs without gaunge invariance}.} (both technically and for the physical interpretation of the amplitudes), or simply group f\/ield theories whenever more general interactions than tensor invariants are allowed (see Section~\ref{sec:simpl} below).

GFTs were originally introduced in the context of Loop Quantum Gravity (LQG) \cite{dPFKR, GFT_rovelli_reisenberg} for the purpose of resumming Spin Foam amplitudes\footnote{Spin foam amplitudes fall in the class of generalized lattice gauge theories which may be generated by a GFT.} \cite{Alexandrov:2011ab, perez_review2012} and hence completing the def\/inition of the dynamics of LQG \cite{rovelli_book, thiemann_book}.
While LQG aims at solving the quantum gravity conundrum through a mere quantization of General Relativity (GR), it naturally leads to quantum state spaces of geometry in which discrete structures such as graphs \cite{ashtekar1995differential} or triangulations \cite{bianca_marc_new} are center stage. Such structures percolate to the dynamical level and lead to interesting quantizations of discretized GR \cite{bo, dl, eprl, fk}, but the question of whether or not a smooth space-time structure can be recovered in some limit remains a great challenge in this approach. In order to address this question, it is in particular crucial to develop new renormalization tools, which should allow to ef\/f\/iciently explore the phase space of spin foam models. Two renormalization programmes have recently emerged to meet this challenge. One is based on an interpretation of spin foam models as direct space regularizations of quantum gravity \cite{bahr2014, bianca_continuum_2014}, and therefore explores generalizations of lattice renormalization techniques. The other one interprets spin foam amplitudes as Feynman contributions in the perturbative expansion of a specif\/ic GFT \cite{lrd_ren, daniele_hydro, tt1}, and therefore requires generalizations of local f\/ield theory renormalization techniques to non-local f\/ield theories such as (T)GFTs. We here review this second research programme, in the particular context of TGFTs with \emph{gauge invariance condition} (or equivalently \emph{closure constraint}), which have been extensively studied in the literature
\cite{frg_r_d,dario_vincentL,thesis,discrete_rg, 4-eps,cor_su2,cor_u1, thomas_reiko, thomas_reiko_sigma,vincentL_constructive, vincentL_constructive2, vincent2_daniele,samary_beta, samary_vignes}. Though not as sophisticated as tentative GFT models for $4d$ quantum gravity \cite{bo_bc, bo}, such theories generate non-trivial spin foam amplitudes and require key generalizations of ordinary renormalization methods. They therefore provide a natural test bed for future and more challenging studies of quantum gravity models.

Our goal is to present the reader with a bird's eye view on the recent literature, and to clearly explain the motivations and the status of the subject. We will as much as possible refrain from delving into technical details. Given the rapid development of the subject, we will also not claim to be exhaustive. The choice of topics to be developed in the main text was at least to some extent a matter of personal taste.

The plan of this review is as follows. In Section~\ref{sec:simpl} we motivate further the GFT renormalization programme, as well as the specif\/ic class of models which has been investigated up to now. Perturbative renormalizability is reviewed in detail in Section~\ref{sec:pert}. A full classif\/ication of renormalizable model based on rigorous power-counting arguments is in particular provided. We then go on to investigations of the properties of the renormalization group f\/lows of these models in Section~\ref{sec:rg}. Emphasis is put on functional renormalization methods, which are of great practical interest even though they have so far only been applied to GFTs in the crudest truncations.

\section{From simplicial to tensorial GFTs}\label{sec:simpl}

\subsection{Renormalization of GFTs: motivations and basic ingredients}

We begin with the introduction of basic GFT structures, which may not be general enough to encompass all models relevant to full $4d$ quantum gravity, but will be suf\/f\/icient in the present context. We def\/ine a GFT as a quantum f\/ield theory for a single complex scalar f\/ield~$\varphi$ leaving on $d$ copies of a f\/ixed compact Lie group~$G$. Unless specif\/ied otherwise, we will use a vector notation for conf\/iguration space variables and its Haar measure
\begin{gather*}
{\bf g} = (g_1, \ldots , g_d) \in G^d, \qquad \mathrm{d} {\bf g} = \mathrm{d} g_1 \cdots \mathrm{d} g_d.
\end{gather*}
We will also use the short-hand notation:
\begin{gather*}
\overline{\varphi}_1 \cdot \varphi_2 = \int \mathrm{d} {\bf g} \overline{\varphi}_1 ( {\bf g} ) \varphi_2 ({\bf g}),
\end{gather*}
for any two square-integrable functions $\varphi_1$ and $\varphi_2$ on $G^d$. The dynamics of the GFT f\/ield is specif\/ied by a probability measure
\begin{gather}\label{measure}
\mathrm{d} \mu_{C_\Lambda}(\varphi , \overline{\varphi}) \exp\left(- S_\Lambda [\varphi, \overline{\varphi}] \right)
\end{gather}
or equivalently a generating functional
\begin{gather}\label{generating}
{\mathcal Z}_\Lambda [J , \bar{J}] = \int \mathrm{d} \mu_{C_\Lambda}(\varphi , \overline{\varphi}) \exp\big({-}S_\Lambda [\varphi, \overline{\varphi}] + \bar{J} \cdot \varphi + \overline{\varphi} \cdot J \big).
\end{gather}
The measure $\mathrm{d} \mu_{C_\Lambda}$ is the Gaussian measure associated to the covariance $C_\Lambda$, which is a positive operator with kernel:
\begin{gather*}
C_\Lambda ( {\bf g} ; {\bf g}' ) = \int \mathrm{d} \mu_{C_\Lambda}(\varphi , \overline{\varphi}) \overline{\varphi}({\bf g}) \varphi({\bf g}').
\end{gather*}
In other words, $C_\Lambda$ is the propagator of the GFT, or equivalently its free $2$-point function. We explicitly introduced an extra regularization or scale parameter $\Lambda > 0$, as the covariance is in general plagued with divergences. The role of this parameter is central in our renormalization programme; we will describe it in greater details once a specif\/ic choice of propagator is made. Perturbations around the Gaussian theory are introduced through the (interaction part of the) GFT action $S_\Lambda[\varphi, \overline{\varphi}]$, which we will parametrize as
\begin{gather*}
S_\Lambda [\varphi, \overline{\varphi}] = \sum_{b \in {\mathcal B}} t_b (\Lambda) I_b [\varphi, \overline{\varphi}],
\end{gather*}
where ${\mathcal B}$ is a set of elementary interactions, $I_b$ is the specif\/ic monomial in the f\/ields associated to the interaction $b$, and $t_b (\Lambda)$ is the running coupling constant associated to~$b$ at scale $\Lambda$. One main objective of renormalization is to determine how these coupling constants should be adjusted to compensate for a change in the cut-of\/f~$\Lambda$. More precisely, as in ordinary f\/ield theory, we will require that the infrared content of the connected $2k$-point functions (or Schwinger functions)
\begin{gather*}
\mathsf{S}^{(2k)}_\Lambda ( {\bf g}_1 , \ldots , {\bf g}_k , {\bf g}_1' , \ldots , {\bf g}_k') = \left. \left( \prod_{i = 1}^{k} \frac{\delta}{\delta \bar{J} ( {\bf g}_i )} \frac{\delta}{\delta J ( {\bf g}_i' )} \right) \ln {\mathcal Z}_\Lambda [J , \bar{J}]\right|_{J= \bar{J} = 0},
\end{gather*}
which also characterize the random measure (\ref{measure}), is invariant under a change of $\Lambda$. Unlike ordinary f\/ield theories, we do not have any prior notion of energy scale which we can rely on to determine what 'infrared' means in the GFT context. We will therefore need to adopt a more abstract notion of scale, {\it{a priori}} unrelated to familiar space-time concepts.

For specif\/ic choices of group, propagator and basic interactions, the formal Feynman expansion of (for instance) the partition function
\begin{gather}\label{spin-foams}
{\mathcal Z}_\Lambda := {\mathcal Z}_\Lambda [0 , 0] = \sum_{\mathcal G} \prod_{b\in {\mathcal B}} \left(- t_b (\Lambda)\right)^{n_b({\mathcal G})} {\mathcal A}_{{\mathcal G}}(\Lambda),
\end{gather}
generates Feynman diagrams ${\mathcal G}$ which are in one-to-one correspondence with specif\/ic (closed) $2$-complexes and are weighted by spin foam amplitudes ${\mathcal A}_{\mathcal G}$\footnote{Note that symmetry factors may have to be included in formula (\ref{spin-foams}), depending on the detailed def\/inition of the model and of the Feynman amplitudes. They are not relevant to the present discussion.}. We remind the reader that, more generally, spin foams are combinatorial objects interpolating between spin-network boundary states, whose amplitudes are constructed as discrete gravity path-integrals taking the form of generalized lattice gauge theory amplitudes. They are therefore interpreted as quantum space-time processes encoding the dynamics of loop quantum gravity spin-network functionals. In this review, we ignore the extra combinatorial structure provided by properly closed boundary spin-network states, since they do not play an essential role in what we want to discuss: the renormalization of such quantum gravity amplitudes is a particular case of that of general $n$-point functions (associated to $n$ open boundary spin-network vertices), we can therefore focus entirely on the latter. In view of equation~(\ref{spin-foams}), the GFT formalism provides natural prescriptions for resumming inf\/inite classes of spin foam amplitudes, with weights parametrized by the GFT coupling constants. This is the sense in which GFTs allow to complete the def\/inition of spin foam models, which cannot be claimed to fully specify a dynamics for quantum gravity unless an extra organization principle for its amplitudes is clearly spelled out. The summing prescription implemented through GFT is one such possible organization principle, which ef\/fectively removes a~large class of discretization ambiguities entering the def\/inition of spin foam models (but not all). What remains to be checked is: 1)~whether this formal procedure is mathematically consistent; and 2)~whether it is physically relevant, in the sense that general relativity can be recovered in some limit. Renormalization will presumably play an important role in order to meet both of these challenges.

The formal relation between GFTs and spin foam models being an intrinsically perturbative statement, checking its validity and consistency is essentially equivalent to proving renorma\-li\-za\-bility of the GFT. At the very least one needs to check that~-- again formally~-- the set of GFT interactions $I_b$ is stable under a shift of the cut-of\/f $\Lambda$. This translates into a formal stability of the set of spin foams summed over on the right-hand side of equation~(\ref{spin-foams}). But turning this rather vague statement into a sensible perturbative def\/inition requires that the same stability holds with only f\/initely many GFT interactions turned on, and that is equivalent to the perturbative renormalizability of the GFT. The relevant GFT interactions will then uniquely determine which (f\/initely many) elementary spin foam interaction vertices dominate in this perturbative phase. This is a more concrete and more rigorous illustration of how the GFT formalism may be powerfully used to remove spin foam discretization ambiguities and make predictions.

Furthermore, assuming that general relativity may only be recovered in a phase in which macroscopically large spin-network boundary states acquire large amplitudes, addressing the second open problem will presumably require to go beyond perturbation theory. This suggests that the GFT phase space should be more systematically explored away from its perturbative regime, and the existence of non-trivial f\/ixed points of the renormalization group investigated. It is important to realize that such non-trivial f\/ixed points would correspond to non-perturbative resummations of spin-foam amplitudes and would therefore be very hard to grasp without recourse to GFT. They will generate new vacua, supporting new and possibly inequivalent representations of the GFT, and therefore leading to new GFT phases and phase transitions. The mechanism of Bose--Einstein condensation has in particular been investigated in this context, leading to interesting reconstructions of smooth homogeneous and spherically symmetric space-time geometries from the GFT formalism, with fascinating applications to cosmology~\cite{gfc_letter, gfc_review, gfc_bounce} and black holes~\cite{gfc_bh}. If such a scenario based on a collective reorganization of the spin foam amplitudes is correct, the f\/ield theory language provided by GFT and the powerful ef\/fective methods it entails seems hardly avoidable.

We now brief\/ly outline some basic features of GFT model-building, the interested reader is referred to reviews on the subject \cite{ad_ten, freidel_gft, Krajewski_rev, daniele_rev2006, daniele_rev2011} and references therein for further details. The specif\/ic GFTs which generate quantum gravity spin foam amplitudes (for instance \cite{bo, Geloun:2010vj, Krajewski:2010yq}) require~$G$ to be a local symmetry group of space-time (or space), e.g., the Lorentz group \mbox{${\rm SO}(1,d-1)$} or its universal covering. In this review, we will only consider Euclidean groups -- in particular ${\rm SU}(2)$ in dimension $d=3$~-- and ignore complications arising from the Lorentzian signature\footnote{The renormalization of GFTs with Lorentzian signature remains largely open and will likely become an active f\/ield of research in the close future. See however~\cite{aldo_eprl} for a f\/irst attempt in this direction.}. Another important ingredient is the so-called \emph{gauge invariance condition}, def\/ined as a global symmetry of the GFT f\/ield under simultaneous translation of its group variables:
\begin{gather}\label{gauge_inv}
\forall\, h \in G, \qquad \varphi( g_1 h, \ldots , g_d h) = \varphi( g_1 , \ldots , g_d ).
\end{gather}
This condition is common to all known proposals of GFT models for quantum gravity, and is responsible for the generalized lattice gauge theory form of the amplitudes appearing on the right-hand side of equation~(\ref{spin-foams}). Within the general GFT formalism spelled out at the beginning of this section, we implement this symmetry by requiring that the covariance~$C_\Lambda$ is of the form
\begin{gather*}
C_\Lambda = {\mathcal P} \tilde{C}_\Lambda {\mathcal P},
\end{gather*}
where $\tilde{C}_\Lambda$ is again a positive operator, and ${\mathcal P}$ is the projector on translation invariant GFT f\/ields with kernel
\begin{gather*}
{\mathcal P}({\bf g} , {\bf g}' ) = \int_G \mathrm{d} h \prod_{\ell=1}^d \delta\big( g_\ell h g_\ell'^{-1} \big).
\end{gather*}
Hence $C_\Lambda$ is degenerate and its image lies within the space of f\/ields verifying (\ref{gauge_inv}). The gauge invariance condition, also called \emph{closure constraint}, is the main dynamical ingredient of GFT models for quantum BF theory in arbitrary dimension. In dimension $3$, it turns out that ${\rm SU}(2)$ BF theory can be interpreted as a theory of Euclidean gravity, and therefore ${\rm SU}(2)$ GFT with closure constraint provides a natural arena in which to formulate $3d$ Euclidean quantum gravity models. A~typical example is the Boulatov model~\cite{boulatov} (which generates Ponzano--Regge spin foam amplitudes), a more recent version of which will be introduced below. In higher dimensions, further conditions on the GFT f\/ields~-- which are known as \emph{simplicity constraints}~-- need to be implemented, possibly leading to further degeneracies of the covariance. In what follows, we will however ignore such constraints and focus on examples in which $\tilde{C}_\Lambda$ is non-degenerate. Finally, in most of the quantum-gravity literature, the $2$-complexes supporting spin foam amplitudes are assumed to be dual to simplicial decompositions of manifolds. This is a~simplif\/ication entering the construction of discrete gravity path-integrals which, though very natural, is as far as we can tell not very well motivated. At the GFT level, this corresponds to a~choice of action~$S_\Lambda$ comprising a unique type of monomials $I_b$, which are of order $(d+1)$ and contract the f\/ield variables pairwise following the pattern of a $d$-simplex. Such models have very rigid combinatorial structure, and therefore their renormalization programme is more dif\/f\/icult to apprehend. Furthermore, radiative corrections which are not of the simplicial type are in general generated by the renormalization group f\/low, and therefore need to be added to~$S_\Lambda$ from the outset. The importance of this simple realization should not be underestimated: in ordinary quantum f\/ield theory, a similar argument implies that one should in principle allow any number of local interactions in the action; it is then left to the renormalization group to identify a f\/inite relevant subset of local operators among all possible interactions. Likewise, a~renormalization programme for GFTs requires the prior identif\/ication of an inf\/inite reservoir of allowed interactions, providing a suitable generalization of ordinary locality.

The purpose of the next three subsections is to explain how recent developments in tensor models made such a generalized notion of locality available and allowed to launch a GFT renormalization programme. We will more specif\/ically focus on GFTs with gauge invariance condition, and explain (including a new heuristic argument presented for the f\/irst time in this review) in which sense so-called \emph{colored GFTs}~\cite{razvan_colors}~-- which are more recent and better behaved versions of the simplicial GFTs brief\/ly mentioned before -- can be embedded in a larger and f\/lexible enough class of models. These theories are known as \emph{tensorial GFTs} and are the main focus of the rest of the review.

\subsection{Simplicial GFT models and tensorial theory space}\label{subsec:simpl}

Colored tensor models \cite{razvan_jimmy_rev} and GFTs were introduced in 2009 by Gurau \cite{razvan_colors} and have since then overcome two important caveats of older simplicial constructions \cite{boulatov, ooguri}: 1) in dimension $d\geq3$, the combinatorial data of Feynman diagrams failed to unambiguously encode the combinatorial structure and topology of the simplicial complexes generated in perturbative expansion; 2)
even though power counting results could be derived \cite{lin_gft, lrd_ren, matteo_scaling}, no consistent organization of the amplitudes -- such as the celebrated $1/N$ expansion of matrix models~-- could be proposed to make sense of the formal perturbative expansion.
We therefore decided to gloss over the original simplicial models and work instead with colored structures from the outset.

\begin{figure}[h]\centering
\includegraphics[scale=1.1]{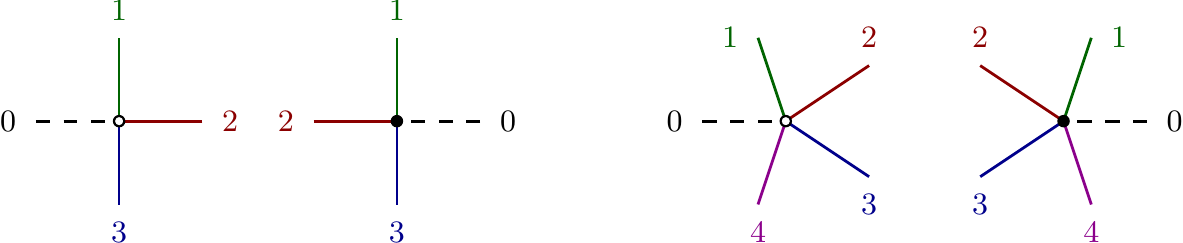}
\caption{Colored interaction vertices in dimension $d=3$ (left) and $d=4$ (right).}\label{colored-v}
\end{figure}

\begin{figure}[h]\centering
\includegraphics[scale=1.1]{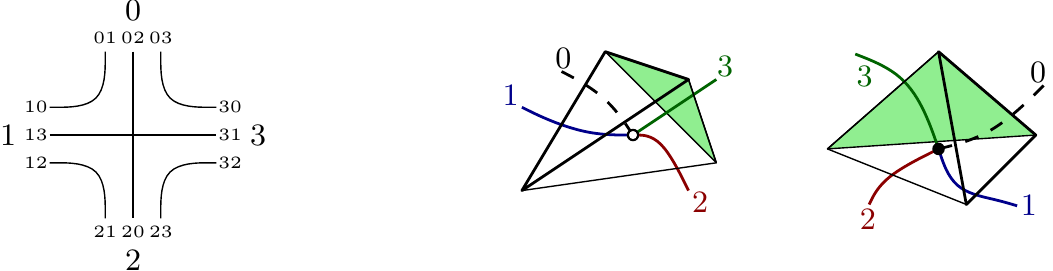}
\caption{Stranded representation of the interactions in dimension $d=3$ (left) and dual simplicial picture (right). Colors of boundary triangles are left implicit, except for green ones.}\label{stranded}
\end{figure}

Within the general GFT set-up we have described, the def\/inition of a colored GFT model in~$d$ dimensions requires the introduction of~$d$ auxiliary complex GFT f\/ields $\{\varphi_c \,\vert\, c = 1, \ldots, d\}$, with covariance $\bar{C}_\Lambda$ not necessarily identical to $C_\Lambda$. The label~$c$ is called \emph{color}, and we conventionally associate the color $0$ to the original f\/ield $\varphi \equiv \varphi_0$. The action $S_\Lambda$ is then implicitly def\/ined by
\begin{gather}\label{colored-uncolored}
\exp\big({-} S_\Lambda [\varphi , \overline{\varphi}] \big) = \int \left[\prod_{c = 1}^d \mathrm{d} \mu_{\bar{C}_\Lambda}(\varphi_c , \overline{\varphi}_c)\right] \exp\big( {-} S^{\mathrm{col}}_\Lambda [\varphi, \overline{\varphi} ; \varphi_c , \overline{\varphi}_c] \big),
\end{gather}
where the colored GFT action is
\begin{gather}
S^{\mathrm{col}}_\Lambda [\varphi_0, \overline{\varphi}_0 ; \varphi_c , \overline{\varphi}_c] = \lambda(\Lambda) \int \left[\prod_{0 \leq \ell < \ell' \leq d} \mathrm{d} g_{\ell \ell'}\right] \prod_{\ell=0}^d \varphi_\ell( {\bf g}_\ell )\nonumber\\
\hphantom{S^{\mathrm{col}}_\Lambda [\varphi_0, \overline{\varphi}_0 ; \varphi_c , \overline{\varphi}_c] =}{}
 + \bar{\lambda}(\Lambda) \int \left[\prod_{0 \leq \ell < \ell' \leq d} \mathrm{d} g_{\ell \ell'}\right] \prod_{\ell=0}^d \overline{\varphi}_\ell( {\bf g}_\ell )\label{action_c}
\end{gather}
and we have used the convention that $g_{\ell \ell'} = g_{\ell' \ell}$ together with the notation
\begin{gather*}
{\bf g}_\ell = (g_{\ell \ell-1},\ldots,g_{\ell 0}, g_{\ell d}, \ldots, g_{\ell \ell+1}).
\end{gather*}
The two interactions are interpreted as pairwise gluings of $(d+1)$ $(d-1)$-simplices along \mbox{$(d-1)$}-subsimplices, following to the shape of a $d$-simplex. This pattern of contractions can be pictorially represented by white (resp.\ black) nodes where $(d+1)$ colored half-edges meet as in Fig.~\ref{colored-v}. An half-edge of color $\ell$ is associated to a GFT f\/ield $\varphi_\ell$ or $\overline{\varphi}_\ell$, and is dual to a~$(d-1)$-simplex of color $\ell$. A pair of edges of colors $\ell$ and $\ell'$ in turn encodes the integral over the va\-riab\-le~$g_{\ell \ell'}$ in formula~(\ref{action_c}); it is interpreted as a pairwise gluing of two dual $(d-1)$-simplices along a~$(d-2)$-subsimplex. In Fig.~\ref{stranded}, we provide an equivalent stranded representation of the pattern of contractions associated to the $3d$ vertices, and illustrate how the dual tetrahedra can be reconstructed from the colored vertices: half-lines are dual to triangles, which are glued pairwise along their boundary edges.

\begin{figure}[t!]\centering
\includegraphics[scale=1]{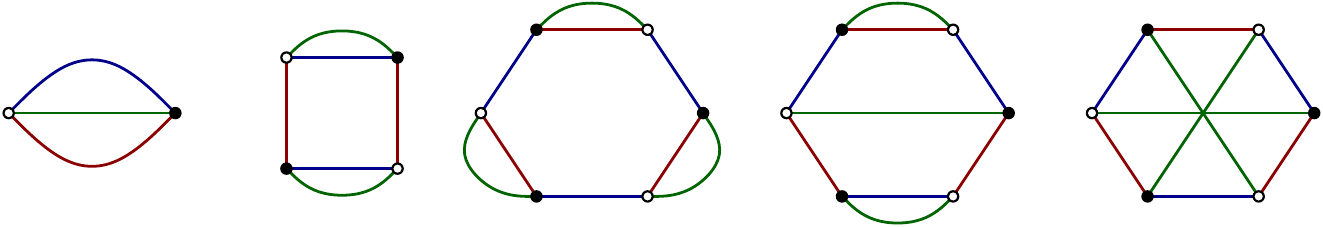}
\caption{Bubble interactions up to order $6$ in $d=3$.}\label{3d-bubbles}
\end{figure}

Note that half-lines associated to f\/ields $\varphi$ and $\overline{\varphi}$ (with color $0$) are dashed. This is to emphasize that the latter are the true dynamical variables of the theory; for instance, in equation~(\ref{generating}) we remark that only them have been coupled to external sources. The colored f\/ields $\varphi_c$ for $c= 1, \ldots , d$ can therefore be (formally) integrated out. This yields ef\/fective interactions $I_b$ parametrized by $d$-colored graphs $b$ involving colors $c=1$ to~$d$. These graphs are also called \emph{bubbles} in the literature (e.g., in~\cite{uncoloring, virasoro}; examples in $d=3$ are provided in Fig.~\ref{3d-bubbles}. The coupling constant $t_b$ is moreover proportional to\footnote{The proportionality factor depends on the combinatorial structure of the bubbles alone, not on the other ingredients of the model.} $(\lambda \bar{\lambda})^{N_b / 2}$, with $N_b$ the number of nodes in the colored graph~$b$.

\begin{figure}[h]
\begin{center}
\includegraphics[scale=1.2]{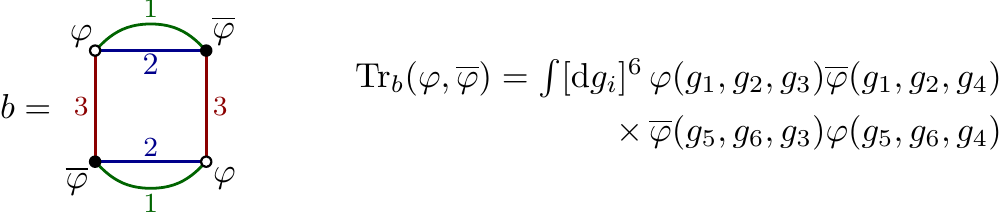}
\caption{A four-valent $3d$ bubble and its corresponding tensor invariant.}\label{invariant_ex}
\end{center}
\end{figure}

\begin{figure}[h]
\begin{center}
\includegraphics[scale=0.8]{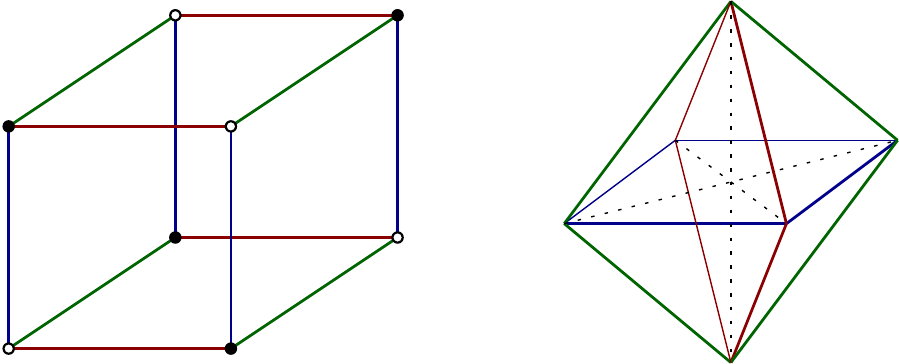}
\caption{A $3d$ bubble of valency $8$ dual to a double pyramid, which can equivalently be viewed as a~gluing of $8$ colored tetrahedra.}\label{3d-complex}
\end{center}
\end{figure}

The precise form of the ef\/fective interaction $I_b$ highly depends on the auxiliary cova\-rian\-ce~$\bar{C}_\Lambda$ and is in general quite involved. Let us discuss the special and simple situation in which
\begin{gather}\label{simplified_propa}
\overline{C}_\Lambda ({\bf g} ; {\bf g}') = \prod_{\ell = 1}^d \delta_\Lambda \big(g_\ell g_\ell'^{{-1}}\big),
\end{gather}
where $\delta_\Lambda$ is a regularized version of the delta function on $G$. More precisely, we assume that $\Lambda$ is a sharp cut-of\/f in the Fourier expansion of $\delta$\footnote{For instance, when $G = {\rm SU}(2)$, one may def\/ine
\begin{gather*}
\delta_\Lambda(g) := \sum_{j \in \frac{\mathbb{N}}{2}\vert j(j+1) \leq \Lambda^2} (2j+1) \chi_j (g),
\end{gather*}
where $\chi_j$ are the characters of ${\rm SU}(2)$.}. Under this condition, it can be shown that the ef\/fective $I_b$ are nothing but \emph{tensor invariants} (up to constant factors and powers of the cut-of\/f $\Lambda$ that we ignore for the moment). We refer the reader to~\cite{virasoro}, in which tensor invariants were f\/irst introduced and where their derivation is described in greater details. Following the literature and by analogy with matrix models, we will use in this case a trace notation $I_b \equiv \operatorname{Tr}_b$. A monomial $\operatorname{Tr}_b (\varphi, \overline{\varphi})$ is uniquely determined by its $d$-colored bubble~$b$, under the following rules:
\begin{itemize}\itemsep=0pt
\item a white (resp.\ black) node of $b$ is associated to a f\/ield $\varphi$ (resp.\ $\overline{\varphi}$);
\item an edge of color $\ell$ represents a convolution of two f\/ield variables, both appearing in the $\ell^{\rm{th}}$ copy of the group $G$.
\end{itemize}
An example is provided in Fig.~\ref{invariant_ex}. In this simplif\/ied context, tensor invariant interactions gene\-ra\-te an inf\/inite-dimensional GFT theory space in which colored simplicial models are embedded as a one-parameter family of models\footnote{Colored simplicial models with covariance~\eqref{simplified_propa} generate coupling constants $t_b$ which are functions of $\lambda \bar\lambda$, hence the one-dimensional character of this subspace of theories.}. Bubble interactions therefore provide a generalized notion of locality of the type we have been arguing for. Topologically, they represent elementary but non-simplicial cells with triangulated boundaries. A suggestive $3d$ example is given in Fig.~\ref{3d-complex}\footnote{Note that the boundary edges of the double pyramid on the right can be canonically colored. This illustrates one of the main advantages of the colored structure: it allows to canonically identify all subsimplices in the complex, and therefore faithfully encode its topology. See, e.g.,~\cite{ferri1986, razvan_jimmy_rev}.}. Note however that bubbles may also be dual to topologically singular\footnote{A topological singularity is def\/ined as a point whose neighbourhood is not homomorphic to a ball. This notion should not be confused with that of a metric singularity.} elementary cells, such as, e.g., a topological cone over a non-spherical $(d-1)$-dimensional manifold. This is precisely the case for the rightmost bubble of Fig.~\ref{3d-bubbles}, which is dual to a topological cone over the $2$-torus.

The relevance of the tensorial theory space for GFT renormalization has been f\/irst pointed out in a seminal paper of Ben Geloun and Rivasseau~\cite{tensor_4d}, who proved renormalizability of a~tensorial GFT without gauge invariance condition. This is a context in which the argument we have just presented is applicable, and the relation between tensor invariant models and colored simplicial ones is therefore clear. The situation is more ambiguous as soon as one introduces gauge invariance or other quantum gravity ingredients. In this case one can make two \emph{a priori} inequivalent choices.
\begin{enumerate}\itemsep=0pt
\item The f\/irst possibility is to choose the covariances $C_\Lambda$ and $\bar{C}_\Lambda$ equal. Both are in particular degenerate, and the ef\/fective interactions $I_b$ become quite complicated and hard to mani\-pu\-late in concrete calculations. From the point of view of known spin foam models, which are derived from simplicial discretizations of formal quantum gravity path-integrals, this is however the most natural construction.
\item The second possibility is to assume, as we have done before, that the auxiliary colored f\/ields $\varphi_c$ have trivial covariance $\bar{C}_\Lambda$. One may argue in this case that imposing suitable spin foam constraints on the remaining dynamical f\/ield $\varphi$ will again lead to legitimate discrete gravity path-integrals, however based on non-simplicial cellular complexes.
\end{enumerate}
Which of these two alternatives is the most appropriate remains an open question, and we will not attempt to resolve it in the present article. We will stick to the second alternative, as GFT renormalization has only been substantially explored in this framework. But before that, we outline an additional heuristic calculation which provides a better grasp of the relation between the two approaches, at least in the context of $3d$ Euclidean quantum gravity.

\subsection[Large $N$ expansion and extended tensorial theory space: heuristic derivation]{Large $\boldsymbol{N}$ expansion and extended tensorial theory space:\\ heuristic derivation}\label{heuristic_tt}

The colored Boulatov model (studied in, e.g., \cite{melonic_phase, diffeos, bubbles, razvan_colors, razvanN}) is a GFT for Euclidean quantum gravity in space-time dimension $d=3$. The group $G$ is therefore taken as the (universal covering) of the local symmetry group of Euclidean space: $G = {\rm SU}(2)$. The propagator with cut-of\/f may be def\/ined as
\begin{gather*}
C_N ( g_1 , g_2 , g_3 ; g_1' , g_2' , g_3' ) = \int_{{\rm SU}(2)} \mathrm{d} h \, \prod_{\ell = 1}^3 K_{1/N^2}\big( g_\ell h g_\ell'^{-1} \big) \underset{N \to + \infty}{\longrightarrow} {\mathcal P}(g_1 , g_2 , g_3 ; g_1' , g_2' , g_3'),
\end{gather*}
where $K_\alpha$ is the heat-kernel on ${\rm SU}(2)$ at time $\alpha$\footnote{This def\/ines a regularization of the delta function in which high spin representations are smoothly cut-of\/f:
\begin{gather*}
K_\alpha (g) = \sum_{j \in \frac{\mathbb{N}}{2}} {\rm e}^{- \alpha j (j+1)}(2j+1) \chi_j (g).
\end{gather*}}, and we denote the cut-of\/f by $N$ instead of~$\Lambda$ in reference to the original literature \cite{razvanN, razvan_complete, RazvanVincentN}. The auxiliary covariances appearing in formula~(\ref{colored-uncolored}) are furthermore taken equal to the propagator: $\bar{C}_N = C_N$. In \cite{razvanN} Gurau showed that, under the assumption that the coupling constant asymptotically behaves like
\begin{gather*}
\lambda(N) \underset{N \to + \infty}{\sim} \frac{\lambda_0}{N^{3/2}}
\end{gather*}
for some f\/ixed $\lambda_0$, the colored Boulatov model admits a $1/N$-expansion. In particular, the partition function can be expanded as
\begin{gather}\label{N}
{\mathcal Z}_N = N^6 {\mathcal Z}_0 (\lambda_0 \bar{\lambda}_0) + N^3 {\mathcal Z}_1 (\lambda_0 \bar{\lambda}_0) + O( 1 ),
\end{gather}
where ${\mathcal Z}_0$ and ${\mathcal Z}_1$ sum over specif\/ic inf\/inite families of Feynman diagrams representing spherical manifolds\footnote{The family summed over by ${\mathcal Z}_0$~-- the melonic graphs~-- has been extensively studied in tensor models (see, e.g., \cite{critical, razvan_jimmy_rev, melon_branched} as well as in the present context~\cite{melonic_phase}. The partition function ${\mathcal Z}_1$ has as far as we know not been studied in great details, but it is nonetheless known that it sums spherical manifolds \cite{razvan_complete}.}. What is of crucial interest for us is that singular topologies, and hence singular ef\/fective interactions, are all convergent. This clear separation between the f\/irst leading contributions in $N$ and the f\/irst topologically singular spin foam structures, already established in~\cite{razvanN}, was more systematically investigated in~\cite{bubbles, ooguri_edge} by means of dif\/ferent methods\footnote{In particular, tighter bounds were derived, showing that singular topologies are suppressed in at least~$N^{3(1-S)}$, where $S$ is the number of singular bubbles. A similar result was shown to hold also in the case of the $4d$ colored Ooguri model~\cite{ooguri_edge}.}. Singular topologies having no natural space-time interpretation at this point\footnote{While metric singularities are generic in general relativity, topological singularities are completely absent of standard models of classical space-time.}, this is a welcomed property of the $1/N$ expansion.

Now, this means that we can truncate the ef\/fective action def\/ined in equation (\ref{colored-uncolored}) to non-singular bubbles without af\/fecting ${\mathcal Z}_0$ and ${\mathcal Z}_1$. By def\/inition, such non-singular bubbles have moreover spherical boundaries, which implies that they lead to bulk amplitudes which are peaked around trivial holonomies. Let us give an illustration of what this means by focusing on the simplest possible bubble: the one with $2$ nodes shown on the leftmost side of Fig.~\ref{3d-bubbles}. It can be shown that it generates a term in the action $S_N$ of the form
\begin{gather*}
I_2 (\varphi, \overline{\varphi}) = \int \mathrm{d} {\bf g} \mathrm{d} {\bf g}' \, \overline{\varphi}({\bf g}) \varphi({\bf g}') \int [\mathrm{d} h_i]^3 K_{1/N^2}\big(h_1 h_2^{-1}\big) K_{1/N^2}\big(h_1 h_3^{-1}\big) K_{1/N^2}\big(h_2 h_3^{-1}\big) \\
\hphantom{I_2 (\varphi, \overline{\varphi}) =}{} \times \prod_{i = 1}^3 K_{1/N^2}\big(g_i h_i g_i'^{-1}\big) .
\end{gather*}
Using the gauge invariant condition (\ref{gauge_inv}), one is free to translate the $g_i$ variables by (say)~$h_3^{-1}$. This reduces the $h_i$ dependence of the last line to a dependence in $h_3^{-1} h_1$ and $h_3^{-1} h_2$. Performing the change of variables $h_1 \to h_3 h_1$ and $h_1 \to h_3 h_2$, we hence obtain an integral which is completely independent of~$h_3$. By normalization of the Haar measure, the new expression of~$I_2$ is thus
\begin{gather}
I_2 (\varphi, \overline{\varphi}) = \int \mathrm{d} {\bf g} \mathrm{d} {\bf g}' \, \overline{\varphi}({\bf g}) \varphi({\bf g}') \int \mathrm{d} h_1 \mathrm{d} h_2 K_{1/N^2}\big(h_1 h_2^{-1}\big) K_{1/N^2}(h_1) K_{1/N^2}(h_2) \nonumber\\
\hphantom{I_2 (\varphi, \overline{\varphi}) =}{} \times K_{1/N^2}\big(g_1 h_1 g_1'^{-1}\big) K_{1/N^2}\big(g_2 h_2 g_2'^{-1}\big) K_{1/N^2}\big(g_3 g_3'^{-1}\big).\label{2-point_fixed}
\end{gather}
This procedure is nothing else than a gauge f\/ixing and is quite general: there is a gauge freedom associated to each node in the bubble, which allows to trivialize the holonomies along a tree of colored edges~\cite{freidel_louapre_PRI}. Now, in the large $N$ limit, one realizes that: the heat-kernels appearing in the f\/irst line of equation~(\ref{2-point_fixed}) render the integrand sharply peaked around $h_1 = h_2 = \one$; together with the second line, this implies that the integrand is also sharply peaked around $g_i = g_i'$. This simple fact entitles us to expand $\varphi({\bf g}')$ in Taylor expansion around ${\bf g}$:
\begin{gather*}
\varphi({\bf g}') = \varphi({\bf g}) + \left.\frac{\mathrm{d}}{\mathrm{d} t}\right|_{t=0}\varphi({\bf g}(t)) + \frac{1}{2} \left.\frac{\mathrm{d}^2}{\mathrm{d} t^2}\right|_{t=0}\varphi({\bf g}(t)) + \cdots,
\end{gather*}
where ${\bf g}'(t)$ is an af\/f\/inely parametrized geodesic from ${\bf g}$ to ${\bf g}'$ in ${\rm SU}(2)^3$. This reduces $I_2$ to an inf\/inite sum over tensor invariant contractions of the f\/ields \emph{and their derivatives}. More precisely, this procedure can only generate ${\rm SU}(2)$-invariant dif\/ferential operators acting on each copy of~${\rm SU}(2)$ and one therefore obtains:
\begin{gather}\label{tensor_laplace}
I_2 (\varphi, \overline{\varphi}) = a(\Lambda) \int \mathrm{d} {\bf g} \, \overline{\varphi}({\bf g}) \varphi({\bf g}) + b(\Lambda) \int \mathrm{d} {\bf g} \, \overline{\varphi}({\bf g}) \left(- \sum_{\ell=1}^3 \Delta_\ell \right) \varphi({\bf g}) + \cdots,
\end{gather}
where $\Delta_\ell$ is the Laplace operator acting on the $\ell^\mathrm{th}$ copy of ${\rm SU}(2)$, and $a(\Lambda)$, $b(\Lambda)$ are computable functions. Higher order terms will involve invariant dif\/ferential operators of arbitrary order.

The previous argument can be applied in full generality. Any ef\/fective vertex $I_b$ associated to a non-singular bubble $b$ can be expanded into tensor invariant contractions of the f\/ields and their derivatives. Moreover, only invariant dif\/ferential operators are allowed in this expansion. By picking up a basis of such operators, one can thus def\/ine a generalized space of bubbles $\overline{{\mathcal B}}\supset {\mathcal B}$ labelling generalized trace invariants $\operatorname{Tr}_{\overline{b}}(\varphi, \overline{\varphi})$. The original bubble interactions $\operatorname{Tr}_{{b}}(\varphi, \overline{\varphi})$ (with $b\in{\mathcal B}$) therefore generate a small subset of generalized tensor invariants, those which do not involve any non-trivial dif\/ferential operator. What our analysis proves is that, up to convergent and topologically singular contributions in the $1/N$ expansion (\ref{N}), the colored Boulatov model generates a one-parameter family of ef\/fective actions in the space of generalized tensor invariants. They therefore provide a suitable GFT theory space for the implementation of the f\/irst strategy proposed at the end of the previous subsection, and also shows that the second approach is actually a truncation of the f\/irst.

Finally, we point out that the same argument can be implemented for the colored Ooguri model, and leads to the def\/inition of generalized tensor invariant interactions for ${\rm Spin}(4)$ in dimension~$4$. It remains however to understand how this heuristic calculation may be gene\-ra\-li\-zed to $4d$ models with simplicity constraints, which have not yet been shown to admit~$1/N$ expansions.

\section{Perturbatively renormalizable TGFTs with closure constraint}\label{sec:pert}

\subsection{A general class of models: local 'potential' approximation}

We are now ready to introduce the class of TGFTs with gauge invariant condition, as def\/ined in the literature. We are still in the general set-up introduced at the beginning of the preceding section. Namely, we consider a complex GFT f\/ield $\varphi$ over $d\geq3$ copies of a compact Lie group~$G$. Its free $2$-point function is assumed to be of the form
\begin{gather}\label{schwing}
C_\Lambda ( {\bf g} ; {\bf g}' ) = \int_{1/\Lambda^2}^{+ \infty} \mathrm{d} \alpha
\int_G \mathrm{d} h \,\prod_{\ell=1}^d K_\alpha \big( g_\ell h g_\ell'^{{-1}} \big),
\end{gather}
which is a regulated version of the formal operator $C = {\mathcal P} \Big( {-} \sum\limits_{\ell = 1}^d \Delta_\ell \Big)^{-1} {\mathcal P}$, $K_\alpha$ and $\Delta_\ell$ being respectively heat-kernels and Laplace operators on~$G$. Equation~(\ref{schwing}) is known as the Schwinger representation of the propagator, and $\alpha$ is accordingly called a Schwinger parameter. As for the interaction action $S_\Lambda$, we assume that it is generated by tensor invariants\footnote{Note that in this review we will always include mass terms in the action rather than the covariance, but the opposite convention is sometimes found in the literature.}:
\begin{gather}\label{tgft_action}
S_\Lambda [\varphi, \overline{\varphi}] = \sum_{b \in {\mathcal B}} t_b (\Lambda) \, \operatorname{Tr}_b [\varphi, \overline{\varphi}].
\end{gather}
Furthermore, since color labels have been introduced as purely auxiliary objects, we will also assume that the action is invariant under permutations of the colors. This implies specif\/ic dependences between some of the coupling constants $t_b$ appearing in equation~(\ref{tgft_action}).

From the point of view of the extended GFT space described at the end of the last section, this is the exact analogue of a local potential approximation in ordinary quantum f\/ield theory. Indeed, locality is here embodied by tensor invariance, which entitles us to call the action~$S_\Lambda$ a local potential\footnote{Even though we cannot def\/ine a potential function, due to the combinatorially non-trivial nature of tensorial locality.}. The only non-local terms are restricted to the f\/irst non-trivial dif\/ferential operators appearing in the Taylor expansion of the most general propagator (see equation~(\ref{tensor_laplace})): the Laplace operators~$\Delta_\ell$.

\begin{figure}[t]\centering
\includegraphics[scale=0.85]{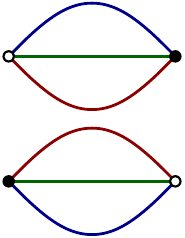}
\caption{A bubble with two connected components.}\label{fig:non-conn}
\end{figure}

Note that in this review we focus more precisely on \emph{connected} tensor invariants. Non-connected bubble contributions may also be included in the action \eqref{tgft_action}, and sometimes cannot be dispensed with. For instance, the diagram shown in Fig.~\ref{fig:non-conn} is associated to the non-connected tensor invariant:
\begin{gather*}
\left( \int \mathrm{d} {\bf g} \, \overline{\varphi}({\bf g}) \varphi({\bf g}) \right)\cdot \left( \int \mathrm{d} {\bf g} \, \overline{\varphi}({\bf g}) \varphi({\bf g}) \right),
\end{gather*}
that is to the product of the tensor invariants encoded by its connected components. Such interactions have for instance appeared in~\cite{samary_vignes}, and have been more systematically studied in~\cite{thomas_reiko}. In the models we will more particularly discuss below, non-connected interactions are irrelevant (as implicitly shown in, e.g.,~\cite{cor_su2}, but more systematically derived in~\cite{thomas_reiko}), and for simplicity we have decided to ignore them altogether.

In order to legitimize the class of TGFTs thus def\/ined as a perfectly honest arena for renorma\-li\-za\-tion, we need to comment a bit more on the notion of scale in this context. For def\/initeness, let us specialize to $G={\rm SU}(2)$, which is one of the most relevant example as far as quantum gravity is concerned. The heat-kernel regularization we have introduced amounts to a smooth regularization of the quantity
\begin{gather*}
p^2 = \sum_{\ell=1}^d j_\ell ( j_\ell + 1),
\end{gather*}
where $\{ j_\ell \}$ are the spin labels associated to the harmonic expansion of the f\/ields. Modes asso\-cia\-ted to $p \geq \Lambda$ are exponentially suppressed, while the theory is essentially untouched at small~$p$. By analogy with ordinary f\/ield theories, we may call $p$ momentum. The fact that we need to regularize large momenta is dictated by the theory itself: this region of the GFT state space is where most of the degrees of freedom lie, and where they produce divergences. Therefore the renormalization group may only f\/low from large to small cut-of\/f\footnote{We remind the reader that the renormalization group is actually not a group: it has a fundamentally directed character since its whole purpose is to erase (irrelevant) physical information.}. Because of that, and by analogy with high energy particle physics, it is conventional in the TGFT literature to dub large (resp.\ small) momenta 'ultraviolet' (resp.\ 'infrared'); we will stick to this nomenclature.

The purpose of our renormalization programme may now be explicitly stated: the goal is to develop the necessary tools for determining the functional dependence of the coupling cons\-tants~$t_b(\Lambda)$, under the condition that the infrared sector of the GFT is kept f\/ixed. We will in particular aim at a complete classif\/ication of perturbatively renormalizable models. This programme is interesting {\it{per se}}, in the sense that it proposes to extend the scope of renormalization theory to quantum f\/ield theories with exotic notions of locality. From a quantum gravity perspective its relevance is on the other hand conditioned by a key conjecture: that it is possible to assume that there is a large separation of scales between the cut-of\/f and the support of interesting $3d$ Euclidean quantum gravity states. It is reassuring to see that this hypothesis is at least superf\/icially consistent. Given that spins label the eigenvalues of the LQG length operator in~$3d$ and that small spins are associated to small lengths, we may for example expect that smooth quantum gravity states can be approximated by (large superpositions of) spin-networks comprising a large number of nodes and edges (or, equivalently, dual triangles and dual edges), but only bounded spins. This also suggests that the study of these smooth quantum gravity states will necessitate a non-perturbative treatment of the GFT renormalization group\footnote{Another related observation is that quantum gravity may require the inclusion of dif\/ferential operators of arbitrary orders, as found in the generalized tensor invariant space evoked previously.}. The present section is devoted entirely to perturbative questions, while some non-perturbative aspects will be discussed in the following one.

\subsection{Power-counting theorem and classif\/ication of models}

We now outline the power-counting arguments leading to the full classif\/ication of renormalizable TGFTs with closure constraint. A very nice feature of TGFTs is that they are amenable to general multiscale methods developed in the context of constructive f\/ield theory~\cite{vincent_book}, allowing rigorous proofs of renormalizability at all orders.

\begin{figure}[t]\centering
\includegraphics[scale=0.85]{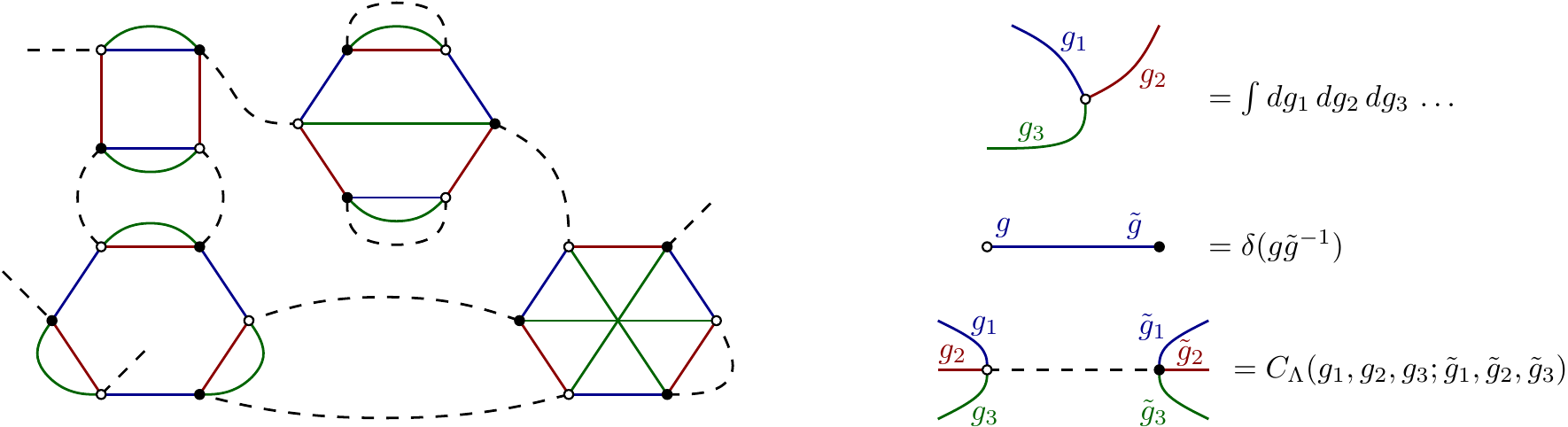}
\caption{Example of TGFT Feynman graph with $N=4$, $V = 4$ and $L = 9$ in $3d$ (left). The amplitude is reconstructed from the drawing by means of the Feynman rules given on the right-hand side.}\label{feynman_rules}
\end{figure}

The Feynman amplitudes of these models are labelled by $(d+1)$-colored graphs in which only dashed (or, equivalently, color-$0$) lines may be open. The $d$-colored connected components without dashed lines, in other words the bubbles, are the interaction vertices, while dashed lines are propagators. Given a Feyman graph ${\mathcal G}$, we will denote by $L({\mathcal G})$, $V({\mathcal G})$ and $N({\mathcal G})$ its set of (internal) dashed lines, bubble vertices and external legs\footnote{As is customary in the literature, we will allow ourselves to use the same notation for the cardinals of these sets.}. Accordingly, the amplitude associated to a graph ${\mathcal G}$ is determined by the following Feynman rules: each node is associated to an integration over $G^d$; each colored line internal to a bubble represents a delta function on~$G$; and f\/inally, dashed lines must be replaced by kernels of $C_\Lambda$. We provide a $3d$ example in Fig.~\ref{feynman_rules}.

The multiscale analysis\footnote{In the present context, see for instance \cite{thesis, cor_su2,cor_u1, samary_vignes}. For a more general and in-depth discussion in local f\/ield theory, \cite{vincent_book} is recommended.} relies on a discrete slicing of the propagator:
\begin{gather}\label{slicing_with_rho}
C_\Lambda = C_{M^{\rho}} = \sum_{i \in \mathbb{N} \vert i\leq \rho} {\mathsf C}_{i},
\end{gather}
where
\begin{gather*}
{\mathsf C}_0 := \int_{1}^{+ \infty} \mathrm{d} \alpha \int_G \mathrm{d} h \prod_{\ell=1}^d K_\alpha \big( g_\ell h g_\ell'^{{-1}} \big), \\
\forall \, i\geq 1, \quad {\mathsf C}_i := \int_{M^{-2i}}^{M^{2(i-1)}} \mathrm{d} \alpha \int_G \mathrm{d} h \prod_{\ell=1}^d K_\alpha \big( g_\ell h g_\ell'^{{-1}} \big).
\end{gather*}
$M > 1$ is a f\/ixed but arbitrary slicing parameter and we have assumed that the UV cut-of\/f is of the form $\Lambda = M^\rho$ with $\rho \in \mathbb{N}$. Each covariance ${\mathsf C}_i$ is then essentially responsible for the propagation of the modes with $M^{i-1} \lesssim p \lesssim M^i$. This procedure induces a decomposition of Feynman amplitudes
\begin{gather}\label{multiscale}
{\mathcal A}_{\mathcal G} = \sum_\mu {\mathcal A}_{{\mathcal G} , \mu},
\end{gather}
according to scale attributions $\mu := \{ i_l \in \mathbb{N} ,\, l \in L({\mathcal G}) \}$. In this formula, ${\mathcal A}_{{\mathcal G}, \mu}$ is an amplitude constructed from the sliced propagators $\{{\mathsf C}_i\}$ rather than the full propagator $C_\Lambda$: that is, to each line $l$, we now associate the propagator ${\mathsf C}_{i_l}$. The multiscale strategy then consists in looking for estimates of each of the amplitudes ${\mathcal A}_{{\mathcal G},\mu}$ separately, rather than of the full amplitude ${\mathcal A}_{\mathcal G}$. The scale attributions $\mu$ have the considerable advantage that they allow to optimize the naive bounds one would derive for ${\mathcal A}_{\mathcal G}$, and as a consequence to more precisely understand the origin of the divergences. For instance, given a graph ${\mathcal G}$ with scale attribution $\mu$, one can realize that divergences may only be generated by \emph{high subgraphs}: these are def\/ined as subgraphs ${\mathcal H} \subset {\mathcal G}$ which have internal scales higher than the scales of their external legs. In order to implement a renormalization procedure, one f\/irst and foremost needs to understand the structure of divergent high subgraphs, and study their behaviour in the limit in which the separation of scales between internal lines and external legs is large.

Without going too much into details, a general Abelian power-counting theorem \cite{lin_gft, cor_su2, samary_vignes} can be derived which, when $G$ is commutative, provides a combinatorial characterization of the divergent subgraphs. When $G$ is non-Abelian, further subtleties enter the picture because, although the Abelian power-counting still holds as a bound \cite{vm1, vm3}, it is not necessarily optimal in this case. However, since it turns out \emph{a~posteriori} that all the divergent graphs encountered in TGFTs do saturate the Abelian bounds\footnote{This is due to the fact that the associated $2$-complexes are simply connected, see again \cite{vm1, vm3}.}, we will simply ignore this subtlety, and the interested reader is referred to \cite{thesis,cor_su2} for more details.

\begin{figure}[t]\centering
\includegraphics[scale=1]{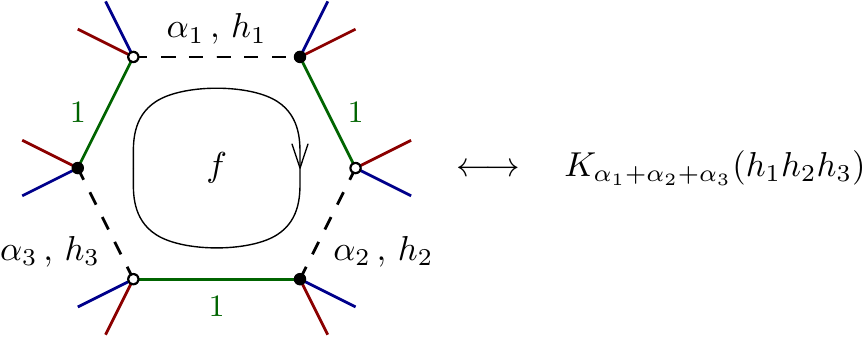}
\caption{An internal face of color $1$ and length $3$, and its associated amplitude integrand.}\label{face}
\end{figure}

Before introducing the notion of degree of divergence, we need to def\/ine the very central notion of \emph{face}. In the present context, a face $f$ of a graph ${\mathcal G}$ is def\/ined as a maximal bicolored path in ${\mathcal G}$, with the restriction that one of the two colors must be $0$. For convenience, we will simply attribute the color $\ell$ to a face consisting of a path with colors $0$ and $\ell$. We will say that a line $l \in L({\mathcal G})$ pertains to $f$ ($l \in f$) if it coincides with one of the dashed lines of the path. Faces can furthermore be \emph{open} (or, equivalently, \emph{external}) or \emph{closed} (or, equivalently, \emph{internal}), depending on whether they are connected to external legs of ${\mathcal G}$ or not. We will denote by $F({\mathcal G})$ (resp.\ $F_{\rm ext} ({\mathcal G})$) the set of closed (resp.\ open) faces of ${\mathcal G}$. Conventionally, we may also orient dashed lines positively from white to black nodes, and orient the faces accordingly. This allows to introduce an adjacency matrix $\varepsilon_{lf}$, of size $L \times F$ and with only $0$ or $1$ entries: $\varepsilon_{lf} = 1$ if $l \in f$, and $\varepsilon_{lf} = 0$ otherwise.
Faces are particularly important because, given the form of the propagator and the Feynman rules, the integrand of an amplitude factorizes over its faces. More precisely, each closed face $f$ yields a factor
\begin{gather*}
K_{\underset{l\in f}{\sum} \alpha_l} \left( \underset{l \in f}{\overrightarrow{\prod}} h_l \right),
\end{gather*}
where $\alpha_l$ and $h_l$ are respectively the Schwinger parameter and holonomy associated to the propagator line~$l$, and the product over holonomies is taken accordingly to the orientation of $f$. See Fig.~\ref{face} for an example. An open face $f$ yields on the other hand a factor
\begin{gather*}
K_{\underset{l\in f}{\sum} \alpha_l} \left( g_{s(f)} \left[ \underset{l \in f}{\overrightarrow{\prod}} h_l \right] g_{t(f)}^{-1} \right),
\end{gather*}
where $g_{s(f)}$ and $g_{t(f)}$ are boundary variables associated to the f\/ields sitting at the source ($s(f)$) and target ($t(f)$) ends of $f$.

\begin{Proposition}
The superficial degree of divergence of a $($non-vacuum$)$ graph ${\mathcal G}$ is
\begin{gather*}
\omega( {\mathcal G} ) := - 2 L({\mathcal G}) + D \left( F({\mathcal G}) - R({\mathcal G}) \right) \geq 0 ,
\end{gather*}
where $R({\mathcal G})$ is the rank of the adjacency matrix $\varepsilon_{lf}$ of ${\mathcal G}$, and $D$ is the dimension of~$G$.
\end{Proposition}

The superf\/icial degree of divergence (which we abbreviate to ``degree of divergence'' or simply ``degree'' in the sequel) captures the UV asymptotic behaviour of the amplitudes. For a single-slice amplitude at scale~$i$ (i.e., $\mu = \{ i_\ell = i ,\, l \in L({\mathcal G})\}$), one can show that ${\mathcal A}_{{\mathcal G}, \mu}$ has an exponential scaling of the form $M^{\omega({\mathcal G})i}$ when $i \to + \infty$\footnote{When $\omega({\mathcal G})=0$ the divergences are logarithmic.}. This corresponds to the situation in which~${\mathcal G}$ contains a single high subgraph~-- itself. The fact that the divergences are in this case essentially controlled by the combinatorial quantity $(F-R)$ was already proven in~\cite{lin_gft}, though in a slightly dif\/ferent context. The analysis of more general scale attributions is based on a step-by-step estimation of the contributions of high subgraphs, from higher to smaller scales, and was f\/irst detailed in~\cite{cor_su2, samary_vignes}. We will not need to go into such details here, which are only relevant for the full rigorous proof of renormalizability. We only point out that the concept of high subgraphs allows a very natural treatment of overlapping divergences, which otherwise lead to somewhat challenging recursive constructions (see~\cite{vincent_book} and references therein).

Once one understands how Feynman amplitudes diverge in the UV, one may try to devise simple criteria of renormalizability, for instance in terms of the dimensions $d$ (space-time) and $D$ (group). To this ef\/fect, it is important to f\/ind a more practical expression of the divergence degree, which puts combinatorial quantities such as the number of external legs $N$ to the forefront. The presence of the rank~$R$, which is a direct consequence of the gauge invariance condition we imposed on the f\/ields, makes it more involved than in TGFTs without this ingredient. Invoking elementary combinatorial relations, one easily proves that\footnote{The variable $\rho$ of this equation is a combinatorial quantity associated to a graph, and has nothing to do with the cut-of\/f appearing in equation~\eqref{slicing_with_rho}.}
\begin{gather}\label{omega_rho}
\omega = D(d-2) - \frac{D(d-2)-2}{2} N + \sum_{k \in \mathbb{N}^*} [(D(d-2)-2) k - D(d-2)] n_{2k} + D \rho,
\end{gather}
where $N$ is the number of external legs and $n_{2k}$ the number of bubbles of valency $k$. The whole non-trivial dependence in the rank has been included in the combinatorial quantity
\begin{gather*}
\rho := F - R - (d-2)(L-V+1) .
\end{gather*}
The key missing ingredient leading to a general classif\/ication of models is a bound on $\rho$. The following proposition, which was f\/irst derived in~\cite{cor_su2}, serves this purpose.
\begin{Proposition}\label{propo_rho}
Let ${\mathcal G}$ be a non-vacuum graph. Then
\begin{gather*}
\rho({\mathcal G}) \leq 0,
\end{gather*}
and $\rho({\mathcal G}) = 0$ if and only if ${\mathcal G}$ is a melonic graph.
\end{Proposition}

\begin{figure}[t]\centering
\includegraphics[scale=1]{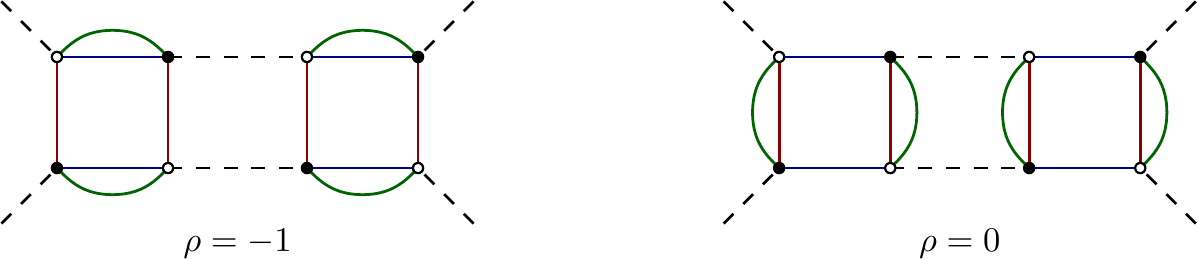}
\caption{Melonic (right) and non-melonic (left) $4$-point graphs in $d=3$. $F=2$ for the former and only $1$ for the latter.}\label{melonic_3d-ex}
\end{figure}

We refer the reader to, e.g., \cite{addendum, cor_su2} for a more precise combinatorial characterization of melonic graphs in this context. As illustrated in Fig.~\ref{melonic_3d-ex}, melonic graphs tend to maximize the number of internal faces. Proposition~\ref{propo_rho} can alternatively be taken as a def\/inition of melonic graphs. We only mention two important properties. First, melonic graphs are associated to and generated by a specif\/ic subset of bubbles, which are accordingly called \emph{melonic bubbles}. In Fig.~\ref{3d-bubbles}, all bubbles are melonic except for the rightmost one, showing that in dimension $d=3$ the f\/irst non-trivial interactions are necessarily melonic. Second, melonic graphs and melonic bubbles have both trivial topology: in dimension $d$ they represent $d$-balls, and are therefore topologically suitable building blocks of $(d+1)$-dimensional space-time.

Proposition \ref{propo_rho} shows that melonic bubbles and melonic graphs lead to the most severe divergences. Following the literature, we now proceed with a classif\/ication of what we may call \emph{melonic models}. We def\/ine them as TGFTs which: 1)~include melonic interactions; and 2)~are perturbatively consistent under renormalization. We emphasize that the f\/irst hypothesis is non-trivial, and is somewhat implicitly assumed in the literature. We will come back to this interesting aspect below.

Power-counting renormalizability requires the degree of divergence to be bounded from above. Moreover, in the presence of divergences, $\omega$ should decrease with the number of external legs, in such a way that only f\/initely many $n$-point functions need to be renormalized. From these conditions alone, one can derive a full classif\/ication of melonic models allowed by the power-counting analysis, in terms of the dimensions $d$ and $D$, and the maximal valency $v_{\max}$ of the renormalizable bubbles. The complete list, established in \cite{cor_su2}\footnote{The classif\/ication of~\cite{cor_su2} includes only just-renormalizable models, we have added super-renormalizable and f\/inite models for completeness.}, is reported in Table~\ref{melonic_models}. Models of type~A to~E are candidate just-renormalizable GFTs, and are in principle the most inte\-res\-ting ones: they have inf\/initely many divergent Feynman amplitudes, which leads to universal properties of the f\/lows. Models of type F and G are on the other hand super-renormalizable, which means that their divergences are generated by a f\/inite family of single-vertex graphs (also known as tadpoles). Finally, models of type H are f\/inite and are therefore not very interesting from the point of view of renormalization: the renormalization group is in this case unable to provide a physical hierarchy for the amplitudes and interactions, which may well all contribute with roughly the same intensity.

\begin{table}[t]\centering
\caption{List of power-counting renormalizable melonic models.}
\label{melonic_models}
\vspace{1mm}
\begin{tabular}{| c || c | c | c |c|c|}
 \hline
 Type & $d$ & $D$ & $v_{{\max}}$ & $\omega$ & Explicit examples\\ \hline\hline
A & 3 & 3 & 6 & $3 - N/ 2 - 2 n_2 - n_4 + 3 \rho$ & $G = {\rm SU}(2)$ \cite{cor_su2, discrete_rg} \\ \hline
B & 3 & 4 & 4 & $4 - N - 2 n_2 + 4 \rho$& $G = {\rm SU}(2)\times {\rm U}(1)$ \cite{4-eps} \\ \hline
C & 4 & 2 & 4 & $4 - N - 2 n_2 + 2 \rho$& -- \\ \hline
D & 5 & 1 & 6 & $3 - N/ 2 - 2 n_2 - n_4 + \rho$& $G = {\rm U}(1)$ \cite{samary_vignes, samary_beta} \\ \hline
E & 6 & 1 & 4 & $4 - N - 2 n_2 + \rho$& $G = {\rm U}(1)$ \cite{samary_vignes, samary_beta, dario_vincentL} \\
 \hline\hline
F & 3 & 2 & arbitrary & $2 - 2 V$ & -- \\ \hline
G & 4 & 1 & arbitrary & $2 - 2 V$ & $G= {\rm U}(1)$ \cite{cor_u1, vincentL_constructive2} \\
\hline\hline
H & 3 & 1 & arbitrary & $1 - L - V < 0$ & $G= {\rm U}(1)$ \cite{vincentL_constructive}\\\hline
 \end{tabular}
\end{table}

A striking feature of this classif\/ication is that the only combination of $d$ and $D$ which is compatible with a quantum space-time interpretation of the amplitudes is $d=D=3$. Indeed, only in this case is the would-be space-time dimension $d$ consistent with the dimension $D$ of the local symmetry group~$G$. This is quite remarkable: we have f\/irst motivated the general class of TGFTs with gauge invariance condition from Euclidean quantum gravity in three dimensions, and reciprocally, pure quantum f\/ield theory arguments allow us to in a sense \emph{derive} dimension~$3$ as the only consistent one.

This prompted an in-depth study of the $d=3$ model with $G={\rm SU}(2)$, which was proven renormalizable at all orders in~\cite{cor_su2}. Its f\/low equations were then studied in greater details in~\cite{discrete_rg}. Note that, because of the subtleties associated to non-Abelian amplitudes, this particular example required extra care, which we are glossing over in this review. Still in $d=3$, a renorma\-li\-zab\-le model of type~B based on the group ${\rm SU}(2)\times {\rm U}(1)$ has been considered in~\cite{4-eps}. Examples of Abelian ${\rm U}(1)$ models of type D and E were proposed in~\cite{samary_vignes} and also proven renormalizable at all orders. Interestingly, and as is clearly allowed by the power-counting arguments we have reviewed, the $\varphi^6$ model of type~D requires the inclusion of a non-melonic and non-connected interaction of the form $(\overline{\varphi} \cdot \varphi)^2$, sometimes called ``anomalous''. The model of type E, which according to our power-counting arguments might also have necessitated the inclusion of non-melonic bubbles\footnote{Graphs with $\rho=-1$ or $-2$ might in principle still lead to divergences.}, remains consistent with only melonic bubbles included\footnote{Note that we are not claiming that non-melonic bubbles cannot be consistently included, only that the model is consistent without them. The construction of non-melonic phases remains a largely unexplored and interesting research direction.}. The beta functions of these two models were then studied in~\cite{samary_beta}, and the functional renormalization group of the model of type~D was investigated in~\cite{dario_vincentL}.

Abelian super-renormalizable models of type G actually provided the f\/irst examples of renormalizable TGFTs with gauge invariance~\cite{cor_u1}. Since only f\/initely many divergent graphs are generated in this case, renormalization could be implemented by means of a generalization of the standard Wick ordering prescription. Constructive aspects of a~$\varphi^4$ model of this type, as well as of a f\/inite model of type H, were more recently studied in \cite{vincentL_constructive, vincentL_constructive2}. This led in both cases to a Borel resummation of the perturbative expansion, thus proving its analytical existence. It is a~very interesting step towards a full non-perturbative def\/inition of just-renormalizable TGFTs with closure constraint, including the more physically relevant $d=3$ and $G={\rm SU}(2)$ situation.

\begin{figure}[t]\centering
\includegraphics[scale=1]{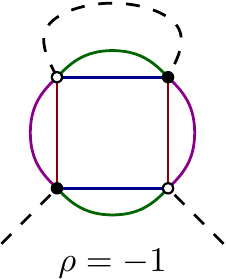}
\caption{Non-melonic $2$-point graph with $\rho = -1$ in $d=4$. Its single vertex is a necklace bubble, which suggests that such an interaction may consistently be included in a TGFT model of type C. The graph has a degree $\omega = 0$ and is therefore logarithmically divergent in this case.}\label{necklace_2point}
\end{figure}

Examples of renormalizable models of type C and F have not been explicitly exhibited in the literature. There is however no doubt that such example exists, for instance with the group $G={\rm U}(1)^2$. Indeed, the arguments and tools from~\cite{cor_u1} and~\cite{samary_vignes} are directly applicable to this Abelian group. In particular, the analysis of non-melonic graphs proposed in~\cite{samary_vignes} allows to demonstrate that melonic bubbles alone lead to a consistent model in situation C. We further conjecture that it is also possible to consistently include~$\varphi^4$ necklace bubbles~\cite{valentin_enhance} in this context, which will generate divergent but non-melonic $2$-point graphs (with $\rho = -1$). See Fig.~\ref{necklace_2point}.

\subsection{Renormalization, subtraction schemes and contractible graphs}\label{tracial_2compl}

Once power-counting renormalizability has been checked, several standard quantum f\/ield theory techniques may be applied to prove full-f\/ledged renormalizability. The main physical idea is to re-express the Feynman expansion in terms of new physically meaningful perturbative parameters. The bare coupling constants are indeed associated to processes occurring at arbitrarily large energies, and have therefore no empirical content. Instead, one should parametrize the theory with the values of $n$-point functions at an arbitrarily chosen but physically accessible low energy scale. In this new expansion, Feynman amplitudes converge, and the divergences of the bare amplitudes are interpreted as spurious ef\/fects resulting from a misplaced parametrization of the f\/ield theory.

Dif\/ferent renormalization prescriptions may be used, leading to slightly dif\/ferent (but equivalent) def\/initions of renormalized quantities. One can for example rely on the celebrated Bogoliubov--Parasiuk--Hepp--Zimmermann (BPHZ) scheme, which amounts to using $p^2 = 0$ as a~reference energy scale. This standard textbook procedure is rather simple at f\/irst loop orders, but may appear somewhat mysterious when it comes to overlapping divergences\footnote{We remind the reader that, given a graph ${\mathcal G}$, two divergent subgraphs ${\mathcal H}_1, {\mathcal H}_2 \in {\mathcal G}$ are said to overlap if neither of the three relations is verif\/ied: ${\mathcal H}_1 \cap {\mathcal H}_2 = \varnothing$, ${\mathcal H}_1 \subset {\mathcal H}_2$, ${\mathcal H}_2 \subset {\mathcal H}_1$.}. The ef\/fective expansion, based on multiscale methods, is more in the spirit of Wilson's brilliant reformulation of renormalization. It consists in a step-by-step recursive def\/inition of ef\/fective coupling constants~$t_{b,i}$, which measure the amplitudes of physical processes associated to the index scale~$i$. The contributions of divergent graphs at a given scale are reabsorbed into the coupling constants at lower scales, resulting in a discrete renormalization group f\/low from higher to lower scales:
\begin{gather*}
t_{b,i-1} - t_{b,i} = \beta_i ( \{ t_{b',j}\,\vert \, i \leq j \leq \rho \} ).
\end{gather*}
The $n$-point functions may then be expressed as formal multi-series in the $t_{b,i}$'s, with convergent coef\/f\/icients in the limit $\rho \to + \infty$. Moreover, the ef\/fective expansion provides a new perspective on the more standard renormalized BPHZ expansion: at f\/ixed multiscale parameter~$\mu$~(\ref {multiscale}), divergent graphs can simply not overlap, and this simple realization greatly clarif\/ies the reason why the BPHZ procedure converges at all orders in perturbation theory. We refer the reader to~\cite{vincent_book} for a detailed discussion of both the BPHZ expansion and the ef\/fective expansion in the context of local scalar f\/ield theories.

Both ef\/fective and renormalized expansions have been successfully generalized and applied to TGFTs with gauge invariance condition~\cite{cor_su2,cor_u1, samary_vignes}. Reviewing these constructions in detail would take us too far into technicalities, we therefore only expose the core argument explaining why these techniques can be applied at all. In ordinary quantum f\/ield theory, renormalization relies on the key realization that high energy processes look essentially local (as seen by an observer operating at much lower energy scales). This is true irrespectively of how complicated these high energy processes are, and is the main reason why the contributions of high divergent subgraphs at a given scale can always be absorbed into redef\/initions of the coupling constants at lower momenta. Possibly severe complications arise in our TGFT context: f\/irst, the non-standard notion of locality encapsulated in tensor invariant interactions renders the ana\-lysis obviously more intricate; second, and more importantly, the main combinatorial building blocks of the amplitudes generated by TGFTs with closure constraint are the faces, which are intrinsically non-local objects.

\begin{figure}[t]\centering
\includegraphics[scale=1]{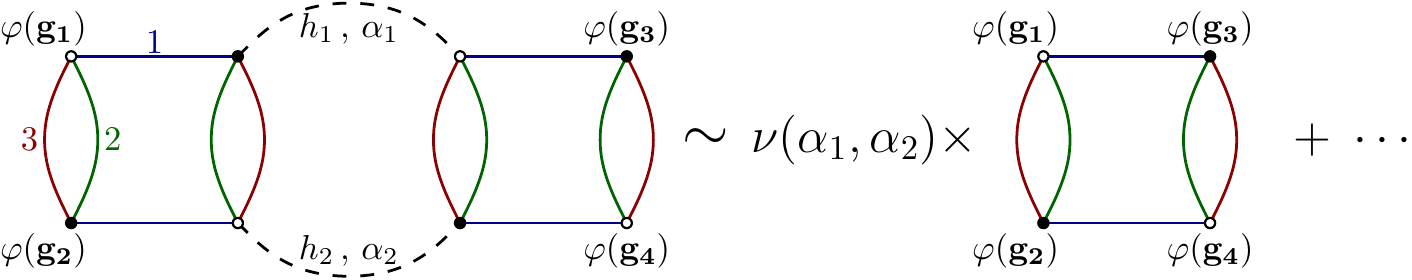}
\caption{Approximation of a tracial graph as an ef\/fective trace invariant contribution plus corrections.}\label{traciality-phi4}
\end{figure}

For concreteness, let us consider the example of Fig.~\ref{traciality-phi4}. On the left one f\/inds a typical ef\/fective contribution to the $4$-point function generated by a melonic graph in $3d$. In terms of the Schwinger parameters~$\alpha_1$ and~$\alpha_2$, the integrand of its amplitude is:
\begin{gather*}
\int \mathrm{d} h_1 \mathrm{d} h_2 \left[ K_{\alpha_1 + \alpha_2} ( h_1 h_2 ) \right]^2 \int \left[\prod_{i<j} \mathrm{d} g_{ij}\right] K_{\alpha_1 } \big( g_{11} h_1 g_{31}^{-1} \big) K_{\alpha_2} \big( g_{21}^{-1} h_2 g_{41} \big) \\
 \qquad {}\times \delta\big( g_{12} g_{22}^{-1} \big) \delta\big( g_{13} g_{22}^{-1} \big)
\delta\big( g_{42} g_{32}^{-1} \big) \delta\big( g_{43} g_{33}^{-1} \big) \varphi( {\bf g}_1 ) \overline{\varphi}( {\bf g}_2 ) \overline{\varphi}( {\bf g}_3 ) \varphi( {\bf g}_4 ).
\end{gather*}
The question is whether this expression can be approximated by an elementary tensor invariant in the sector $\alpha_1, \alpha_2 \to 0$\footnote{We remind the reader that the Schwinger parameters should be thought of as inverse squared momenta~$p^{-2}$.}. Though it is not that obvious at f\/irst sight, the answer is yes. We can resort to a similar line of arguments as the one which led us to the expansion~(\ref{tensor_laplace}). The gauge symmetry associated to the amplitudes allows to gauge-f\/ix one of the two holonomies and reduce this expression to:
\begin{gather*}
\int \mathrm{d} h \left[ K_{\alpha_1 + \alpha_2} ( h ) \right]^2 \int \left[\prod_{i<j} \mathrm{d} g_{ij}\right] K_{\alpha_1 } \big( g_{11} h g_{31}^{-1} \big) K_{\alpha_2} \big( g_{21}^{-1} g_{41} \big)\\
 \qquad {}\times \delta\big( g_{12} g_{22}^{-1} \big) \delta\big( g_{13} g_{22}^{-1} \big)
\delta\big( g_{42} g_{32}^{-1} \big) \delta\big( g_{43} g_{33}^{-1} \big) \varphi( {\bf g}_1 ) \overline{\varphi}( {\bf g}_2 ) \overline{\varphi}( {\bf g}_3 ) \varphi( {\bf g}_4 ).
\end{gather*}
It is then obvious that, in the large scale limit, $h$ is peaked around the identity and therefore $g_{11}$ (resp.~$g_{21}$) is identif\/ied to~$g_{31}$ (resp.~$g_{41}$). A Taylor expansion of the external variables of color $1$ along their external faces then allows to approximate the amplitude by the tensor invariant (and melonic) interaction shown on the right-hand side of Fig.~\ref{traciality-phi4}, up to a scale-dependent constant $\nu(\alpha_1, \alpha_2)$.

We say that a graph is \emph{contractible} if its bulk holonomies can be trivialized in the UV region, as illustrated in the example before\footnote{This is equivalent to simple-connectedness of the $2$-complex, i.e., f\/lat connections are trivial up to gauge.}. Only if all divergent graphs are contractible can we reabsorb UV divergences into tensor invariant ef\/fective interactions. It may however happen that disconnected ef\/fective bubbles are generated in this way. A contractible graph generating a~connected tensor invariant interaction is furthermore called \emph{tracial}. We now understand that the reasons why renormalization theory meaningfully applies to all the candidate models we have introduced in the previous subsection are that: 1)~melonic graphs are tracial; and 2)~all non-melonic divergent corrections are contractible. It is important to understand that these are highly non-trivial facts, which rely on intimate relations between the topology of colored graphs and the scaling of their amplitudes. On the one hand, trivial topology in the bulk of divergent graphs ensures that our tensor invariant truncation is stable under renormalization. But there is more as, in return, this consistent renormalization scheme guarantees that no topological singularities can be generated by radiative corrections. For instance, the rightmost interaction of Fig.~\ref{3d-bubbles} can be consistently set to~$0$ in the $d=3$ model on~${\rm SU}(2)$, even though it is allowed by our power-counting arguments~\cite{cor_su2}. Interestingly, divergent non-melonic graphs with non-melonic bubbles such as the one shown in Fig.~\ref{necklace_2point} are also tracial, which suggests again that necklace terms may sometimes be consistently included into the picture.

\section{Renormalization group and non-trivial f\/ixed points}\label{sec:rg}

The purpose of this section is to illustrate some of the interesting properties of the renor\-ma\-li\-zation group f\/lows of TGFTs with gauge invariance condition. The goal of renormalization group investigations is two-fold: 1) understand the fate of the renormalized coupling constants in the deep UV, and hence determine whether the theory is consistent to arbitrary high scales or not; 2) systematically explore the theory space away from the perturbative f\/ixed point, and investigate in particular the existence of non-trivial f\/ixed points. The f\/irst objective can be f\/irst addressed within a perturbative scheme, and is embodied in this context by the question of \emph{asymptotic freedom}. In the unfavourable case in which asymptotic freedom does not hold (which means that the renormalized coupling constants do not converge to $0$ in the UV), there is still the possibility that the theory may be UV completed by means of a non-perturbative UV f\/ixed point. The second objective is central to the whole GFT approach to quantum gravity and requires a non-perturbative treatment anyway. We therefore decide to focus on the Functional Renormalization Group (FRG), and more precisely on the Wetterich equation, which can conveniently be used to address both types of questions and will presumably play an important role in the future. The related Polchinski equation was on the other hand investigated in \cite{polch_gft, thomas_reiko, thomas_reiko_sigma}.

\subsection{Ef\/fective average action and Wetterich equation}

Irrespectively of one's preferred formulation of the renormalization group, consistent f\/low equations may only be formulated for \emph{dimensionless} coupling constants $u_b (\Lambda)$, which are appropriate rescalings of $t_b(\Lambda)$ by powers of~$\Lambda$. This important aspect may be formalized by the notion of \emph{canonical dimension} $d_b$ of tensor invariant bubbles, which has been discussed at length in~\cite{dario_vincentL,discrete_rg, thomas_reiko}. For melonic models (with closure constraint), it is def\/ined as
\begin{gather*}
d_b := D(d-2) - ( D(d-2) - 2) \frac{N_b}{2},
\end{gather*}
where $N_b$ is the valency of the bubble $b$ (i.e., its number of nodes). Hence the dimensionless coupling constants are def\/ined as
\begin{gather*}
u_b (\Lambda) := \frac{t_b (\Lambda)}{\Lambda^{d_b}},
\end{gather*}
and $b$ is renormalizable (or, equivalently, perturbatively relevant) if and only if $d_b \geq 0$. In view of the expression~(\ref{omega_rho}) for the divergence degree $\omega$, it appears that the most divergent graphs contributing to the running of~$t_b$ diverge like~$\Lambda^{d_b}$, and yield corrections to the dimensionless couplings~$u_b$ of order~$1$ (as they should).

We insist on the fact that the notion of canonical dimension just def\/ined is \emph{well-suited to melonic models only}, as we have implicitly assumed that the most divergent contributions come from melonic graphs. A model which would for instance contain only necklace interactions would bring us out of the melonic world described in this review, and henceforth yield a dif\/ferent notion of canonical dimension\footnote{This possibility is actively explored in the context of $4d$ models with Barrett--Crane simplicity const\-raints~[Carrozza S., Lahoche V., Oriti D., work {i}n progress].}.

\begin{figure}[t]\centering
\includegraphics[scale=1]{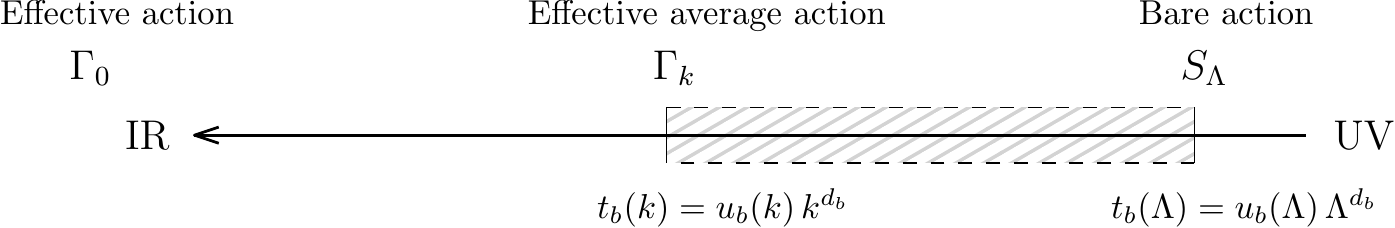}
\caption{The ef\/fective average action interpolates between the bare action in the ultraviolet and the full ef\/fective action in the infrared.}\label{effective_action}
\end{figure}

In the Wetterich--Morris \cite{morris, wetterich_eq} approach to the functional renormalization group, a one-parameter family of deformed generating functionals is introduced
\begin{gather*}
\mathcal{Z}_{\Lambda, k}[J,\bar{J}]:=\int \mathrm{d}\mu_{C_\Lambda}(\varphi,\overline{\varphi}) \exp\big( {-}S_{\Lambda}[\varphi,\overline{\varphi}]-\overline{\varphi} \cdot R_{k} \cdot \varphi + \bar{J} \cdot \varphi + \overline{\varphi} \cdot J \big).
\end{gather*}
The new operator $R_k$ has the function of regularizing the f\/ield modes below an infrared cut-of\/f $k$ while leaving the high energy sector unaf\/fected. This allows the introduction of a new generating functional, the \emph{effective average action} $\Gamma_{k}$, which is the Legendre transform of $W_{k} [J,\bar{J}]:=\ln(\mathcal{Z}_{k, \Lambda} [J,\bar{J}])$, appropriately shifted by the $2$-point counter-term $\overline{\varphi} \cdot R_{k} \cdot \varphi$ (see~\cite{dario_vincentL} for a detailed discussion). Interestingly, for suitable choices of $R_k$, $\Gamma_k$ can be shown to interpolate between the bare action $S_\Lambda = \Gamma_\Lambda$ in the UV, and the full ef\/fective action $\Gamma_0$ (or in other words the generating functional of one-particle irreducible graphs) in the infrared, which we illustrate in Fig.~\ref{effective_action}.

Finally, as derived in detail in \cite{dario_vincentL} for TGFTs with gauge invariant condition, the ef\/fective average action verif\/ies a Wetterich equation
\begin{gather}\label{Wettericheq}
k\partial_k \Gamma_k [\varphi, \overline{\varphi}]=\int \mathrm{d}\textbf{g}_1 \mathrm{d}\textbf{g}_2 \mathrm{d}\textbf{g}_3 k\partial_k R_k (\textbf{g}_1 ; \textbf{g}_2) \big(\Gamma_k^{(2)}+R_k\big)^{-1}(\textbf{g}_2,\textbf{g}_3 ) {\mathcal P}(\textbf{g}_3;\textbf{g}_1),
\end{gather}
where
\begin{gather*}
\Gamma^{(2)}_k [\varphi, \overline{\varphi}]({\bf g} ; {\bf g}'):=\frac{\delta^2 \Gamma_k}{\delta\varphi({\bf g}) \delta\overline{\varphi} ({\bf g}')} [\varphi, \overline{\varphi}]
\end{gather*}
is the full interacting propagator at scale $k$ and ${\mathcal P}$ is as before the projector onto gauge inva\-riant f\/ields. Equation~(\ref{Wettericheq}) def\/ines a formal f\/low in an inf\/inite-dimensional space of theories, a rigorous mathematical def\/inition of which is for the time being out of reach. The standard procedure used in the literature to make sense of the Wetterich equation consists in choosing a~f\/inite-dimensional ansatz for the ef\/fective average action, and then systematically projecting the formal f\/low equation down to this f\/inite-dimensional subspace of theories. The local potential approximation introduced in the previous section is one such possible ansatz, and yields a system of one-loop f\/low equations. Standard one-loop perturbative equations may be recovered with a truncation which only includes renormalizable interactions, while the computation of higher order loops requires the inclusion of non-renormalizable corrections to the potential. The same method can be used in the non-perturbative regime, with the important caveat that truncations are much harder to justify in this case. The functional renormalization group equation provides little analytical control over error terms, and one must therefore resort to more empirical justif\/ications, based mainly on numerical tests of convergence of the truncation procedure. It turns out at the end of the day that the FRG provides reliable and ef\/fective methods for discovering and computing the properties of non-trivial f\/ixed points in ordinary statistical f\/ield theories (see, e.g., \cite{review_frg}). This is what makes them particularly precious in GFTs, as they have the potential to unravel new and more physical phases. Applications of the FRG to TGFTs being rather recent, only the simplest truncations have been considered so far, and the question of their reliability remains to be further explored.

\subsection{Example: perturbative treatment of rank-3 models}

Let us start with a perturbative application of the formalism just introduced, which already suggests a variety of dif\/ferent properties for the renormalization group f\/lows of TGFTs. We restrict our attention to $d=3$ renormalizable models of the type~A ($D=3$) and B ($D=4$). Interestingly, the situation is reminiscent of that of an ordinary local scalar f\/ield, which is renormalizable up to quartic interactions in space-time dimension $4$, and up to order $6$ interactions in dimension~$3$. The same statement holds for $d=3$ melonic models once the space-time dimension is traded for the group dimension~$D$. This remarkable fact allows a simple but informative comparison of the qualitative features of the f\/low equations of TGFTs against those of ordinary local f\/ield theories.

Let us start with a model of type $B$ and choose for instance the group ${\rm SU}(2)\times{\rm U}(1)$ as in~\cite{4-eps}. This theory is renormalizable up to order $4$, therefore a natural ansatz for the ef\/fective average action is\footnote{We use the short-hand notation $\Delta:= \sum\limits_{\ell=1}^3 \Delta_\ell$.}
\begin{gather*}
\Gamma_k (\varphi , \overline{\varphi}) = - Z(k) \, \overline{\varphi} \cdot \Delta \varphi + Z(k) u_{2}(k) k^2 \; \vcenter{\hbox{\includegraphics[scale=0.6]{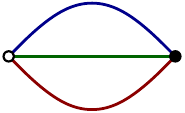}}} + Z(k)^2 u_{4} (k) \; \vcenter{\hbox{\includegraphics[scale=0.6]{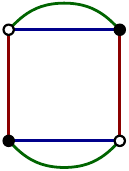}}}\;,
\end{gather*}
where for convenience we directly represented tensor interactions by their associated colored graphs\footnote{Note that each colored graph in this equation is to be thought of as representing the equivalent class of bubbles with the same combinatorial structure up to permutation of the color labels. There is for instance a sum of three distinct melonic interactions at order $4$, corresponding to three inequivalent permutations of the color labels.}, and we have introduced a wave-function parameter~$Z(k)$. In this group dimension, the mass term has canonical dimension $2$ and the marginal $\varphi^4$ interactions have as they should dimension~$0$. It was shown in~\cite{4-eps} that the perturbative renormalization group f\/low reduces in this truncation to
\begin{gather}
k \frac{\partial u_{2}(k)}{\partial k}= - 2 u_{2}(k) - 3 \pi u_{4}(k) ,\nonumber\\
k \frac{\partial u_{4}(k)}{\partial k} = - 2 \pi {u_{4}(k)}^2 .\label{rg_4d}
\end{gather}
One notices a major dif\/ference with ordinary scalar f\/ield theories: the derivative of the $\varphi^4$ coupling is negative, which means that it decreases towards $0$ in the ultraviolet. We therefore obtain an asymptotically free and UV complete perturbative def\/inition of the theory! This is due to a quite general and remarkable property of TGFTs at large. As was f\/irst remarked in~\cite{josephaf} in the context of TGFTs without gauge invariance condition, and later on generalized to models with closure constraint in~\cite{samary_beta}, wave-function counter-terms generally dominate over vertex renormalization ones, and are ultimately responsible for changes in the signs of some of the coef\/f\/icients of the f\/low equations. A~beautiful explanation based on a symmetry argument has been furthermore proposed in~\cite{vincent_af}, thereby proving that asymptotic freedom is a completely general feature of quartic renormalizable TGFTs. In particular, Abelian models of type E are asymptotically free, as argued for in~\cite{samary_beta}.

Let us now move on to the $d=3$ model with $G = {\rm SU}(2)$, which is renormalizable up to order $6$ interactions. Accordingly, one can choose the following ansatz for the ef\/fective average action:
\begin{gather*}
\Gamma_k (\varphi , \overline{\varphi}) = - Z(k) \overline{\varphi} \cdot \Delta \varphi + Z(k) u_{2}(k) k^2 \; \vcenter{\hbox{\includegraphics[scale=0.6]{int2}}} + Z(k)^2 u_{4} (k) k \; \vcenter{\hbox{\includegraphics[scale=0.6]{int4}}} \\
\hphantom{\Gamma_k (\varphi , \overline{\varphi}) =}{} + Z(k)^3 u_{6,1}(k) \; \vcenter{\hbox{\includegraphics[scale=0.6]{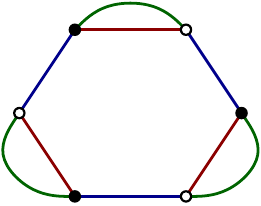}}} + Z(k)^3 u_{6,2}(k) \; \vcenter{\hbox{\includegraphics[scale=0.6]{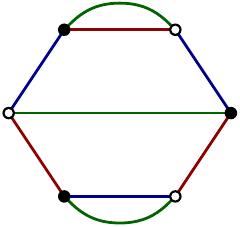}}} \;,
\end{gather*}
and we note the change of dimensionality of the $\varphi^4$ interactions with respect to the previous situation. The one-loop perturbative f\/low can in this case be approximated by\footnote{These f\/low equations were not explicitly evaluated in~\cite{4-eps}, but they immediately follow from the computations reported there.}
\begin{gather}
k \frac{\partial u_{2}(k)}{\partial k} \approx - 2 u_{2}(k) - a u_{4}(k) ,\nonumber\\
k \frac{\partial u_{4}(k)}{\partial k} \approx - u_{4}(k) - b ( u_{6,1}(k) + 2 u_{6,2}(k)) ,\nonumber \\
k \frac{\partial u_{6,1}(k)}{\partial k} \approx - c u_{4}(k) u_{6,1}(k),\nonumber\\
k \frac{\partial u_{6,2}(k)}{\partial k} \approx - d u_{4}(k) u_{6,2}(k),\label{rg_3d}
\end{gather}
where $a$, $b$, $c$ and $d$ are strictly positive constants.
This is a rather complicated system of equations, but we notice again the negative signs in the last two equations. It is therefore tempting to conjecture that, if one assumes that all the coupling constants are positive, then the marginal constants~$u_{6,1}$ and~$u_{6,2}$ both converge to $0$ in the ultraviolet. As shown in~\cite{discrete_rg}\footnote{This paper relied on dif\/ferent methods, in the language of the multiscale expansion. It also went further in that $2$-loop contributions were included to account for quadratic terms (in $u_{6,1}^2$, $u_{6,1}^2$ and $u_{6,1} u_{6,2}$) which have been neglected in (\ref{rg_3d}). The main conclusions are however the same.}, this is actually misleading because, given the form of the f\/low equations, it can be proven that~$u_4 (k)$ necessarily reaches negative values for large $k$, which has the ef\/fect of making the~$\varphi^6$ interactions grow again (see~\cite{discrete_rg}). Hence, one is forced to conclude that this model \emph{cannot} be asymptotically free if we assume that both positive marginal coupling constants are positive\footnote{The situation is a more subtle when $u_{6,1} u_{6,2}<0$, we do not exclude the possibility that one could def\/ine an asymptotically free theory with positive action in this particular case.}. This example shows that the question of asymptotic freedom can be tricky in~$\varphi^6$ renormalizable models, because the~$\varphi^4$ super-renormalizable coupling constants have a non-trivial inf\/luence on the marginal~$\varphi^6$ interactions. In particular, the combinatorial TGFT of~\cite{josephaf} and the type~$D$ model of~\cite{samary_beta}, which have both been argued to be asymptotically free on the basis of an analysis which neglected the f\/low of super-renormalizable constants, may possibly suf\/fer from a similar back-reaction ef\/fect.

\begin{figure}[t]\centering
\includegraphics[scale=.8]{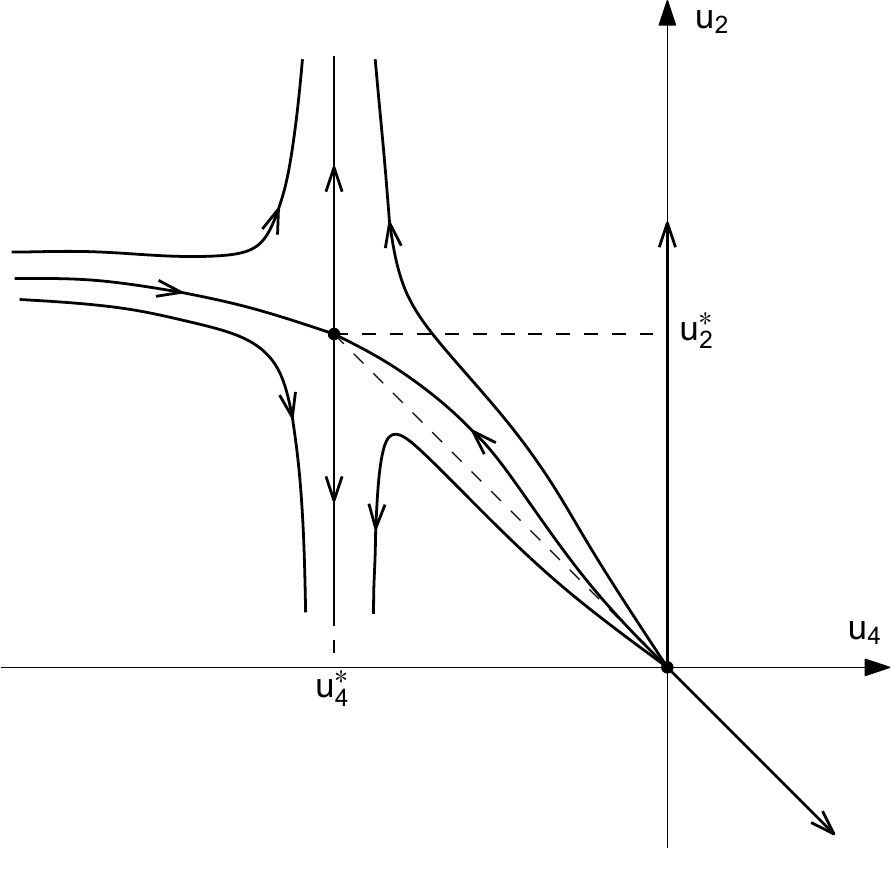}
\caption{Qualitative phase portrait of the renormalization group f\/low equations (\ref{rg_4-eps}), with arrows oriented from high to small scales.}\label{portrait}
\end{figure}

As mentioned already, if a model is not asymptotically free, one may still contemplate the idea of f\/inding a non-perturbation UV completion of it. This is an interesting but notoriously hard question to investigate, since this requires to establish the existence of a non-perturbative f\/ixed point of the renormalization group. An elementary standard method often invoked in statistical physics to test this assumption is the $\varepsilon$-expansion \cite{wilson1974renormalization}, which has the advantage of being essentially perturbative. In scalar f\/ield theories for instance, one can formally def\/ine statistical models in dimension $4- \varepsilon$, which smoothly interpolate between dimension~$4$ and dimension~$6$. A TGFT generalization of this construction was proposed in \cite{4-eps}. The procedure consists in def\/ining a $d=3$ TGFT on the group ${\rm SU}(2) \times {\rm U}(1)^{D-3}$ for arbitrary $D \geq 3$, and then analytically continue the parameter $D := 4 - \varepsilon$ to the interval $3 \leq D \leq 4$. When $\varepsilon$ is small enough, one may assume that the $\varphi^4$ truncation remains pertinent:
\begin{gather*}
\Gamma_k (\varphi , \overline{\varphi}) = - Z(k) \overline{\varphi} \cdot \Delta \varphi + Z(k) u_{2}(k) k^2 \; \vcenter{\hbox{\includegraphics[scale=0.6]{int2}}} + Z(k)^2 u_{4} (k) k^\varepsilon \; \vcenter{\hbox{\includegraphics[scale=0.6]{int4}}}\;.
\end{gather*}
Note that $u_4$ acquired a small canonical dimension $\varepsilon$. This has the ef\/fect of slightly modifying the f\/low equations (\ref{rg_4d}) to
\begin{gather}
k \frac{\partial u_{2}(k)}{\partial k} \approx - 2 u_{2}(k) - 3 \pi u_{4}(k) ,\qquad
k \frac{\partial u_{4}(k)}{\partial k} \approx - \varepsilon u_4 (k ) - 2 \pi {u_{4}(k)}^2 .\label{rg_4-eps}
\end{gather}
Accordingly, one formally f\/inds a new solution to the f\/ixed point equation:
\begin{gather*}
u_{2}^* \approx \frac{3}{4} \varepsilon + {\mathcal O}\big(\varepsilon^2\big) ,\qquad u_{4}^* \approx - \frac{1}{2 \pi} \varepsilon + {\mathcal O}\big(\varepsilon^2\big).
\end{gather*}
Qualitatively, the renormalization group f\/low (\ref{rg_4-eps}) is therefore as represented in Fig.~\ref{portrait}. Extrapolating to $\varepsilon = 1$, this suggests the existence of a TGFT analogue of the Wilson--Fisher f\/ixed point of $3d$ local scalar f\/ield theory in the ${\rm SU}(2)$ model of \cite{cor_su2}. This hypothesis should of course be taken with a grain of salt, since the f\/irst terms in the $\varepsilon$-expansion may only give a~crude idea of what is really going on at $\varepsilon = 1$. Note also that $u_4^*$ has the `wrong' sign (again because wave-function counter-terms dominate in TGFT), which might be taken as a~sign that the formal f\/ixed point we found is only a spurious ef\/fect. This anyway provides solid motivations for performing a~non-perturbative study of the Wetterich equation directly in the case $\varepsilon = 1$ [Carrozza S., Lahoche V., work {i}n progress].

\subsection{Non-perturbative aspects and truncations}

As far as TGFTs with gauge invariance are concerned, the non-perturbative aspects of the Wetterich equation have been studied in two complementary papers \cite{frg_r_d, dario_vincentL}.

In \cite{dario_vincentL}, the role of gauge invariance was carefully analyzed and the Wetterich equation~(\ref{Wettericheq}) was formally derived. The $d=6$ melonic model on ${\rm U}(1)$ (type E) was studied in the $\varphi^4$ melonic truncation
\begin{gather*}
\Gamma_k (\varphi , \overline{\varphi}) = - Z(k) \overline{\varphi} \cdot \Delta \varphi + Z(k) u_{2}(k) k^2 \; \vcenter{\hbox{\includegraphics[scale=0.7]{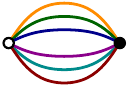}}} + Z(k)^2 u_{4} (k) \; \vcenter{\hbox{\includegraphics[scale=0.7]{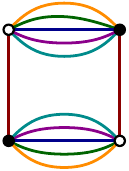}}}\;.
\end{gather*}
For such Abelian models, it is convenient to use a Litim cut-of\/f operator~$R_k$~\cite{litim} which, in momentum space, is def\/ined by the kernel
\begin{gather*}
R_k ({\mathbf{p}}; {\mathbf{p}}) = Z(k) \big( k^2 - {\mathbf{p}}^2\big) \Theta(k^2 - {\mathbf{p}}^2) \prod_{\ell = 1}^6 \delta_{p_\ell p_\ell'},
\end{gather*}
where $p_\ell \in \mathbb{Z}$ label ${\rm U}(1)$ representations. The merit of this type of cut-of\/f is that it greatly simplif\/ies the structure of the truncated f\/low equations, leading to beta functions which are algebraic fractions in the coupling constants. In the perturbative regime one can check again than the model is asymptotically free, as f\/irst proven in~\cite{samary_beta} with dif\/ferent methods. An important aspect of TGFTs on compact Lie groups is that, because of f\/inite size ef\/fects, the renormalization group f\/low is not autonomous. The beta functions of the dimensionless coupling constants explicitly depend on the infrared cut-of\/f $k$, which complicates the search for non-trivial f\/ixed points. The UV ($k\gg 1$) and IR ($k\to 0$) asymptotic regimes were analyzed separately in~\cite{dario_vincentL}. One f\/inds in both a non-perturbative f\/ixed point analogous to the Wilson--Fisher f\/ixed point of statistical f\/ield theory. The quantitative values of the coupling constants are slightly dif\/ferent in the two regimes, but the qualitative structure of the phase portrait is the same. In particular, and unlike the formal f\/ixed point found in the $\varepsilon$-expansion of $3d$ models, the coupling constant~$u_4$ at the f\/ixed point is positive (and~$u_2$ is negative).

A non-compact version of this model, based on the Abelian group $\mathbb{R}$, was investigated in~\cite{frg_r_d}. The renormalization group analysis requires in this case a further regularization of inf\/inite volume divergences. This was implemented by a compactif\/ication of $\mathbb{R}$ into ${\rm U}(1)$, thus resulting in a regularized theory identical to~\cite{dario_vincentL}. The authors could then def\/ine an appropriate thermodynamic limit, capturing the properties of the model in the limit in which the volume of~${\rm U}(1)$ is scaled to inf\/inity. Once more, a f\/ixed point of the Wilson--Fisher type is found in this truncation, consistently with the results of~\cite{dario_vincentL}. It remains to be seen whether this qualitative behaviour will survive closer scrutiny. One will in particular need to check its regularization independence and its stability under extensions of the truncation. This preliminary result may nonetheless be interpreted as a hint of a phase transition between a symmetric and a broken phase of a condensate type, in which the f\/ield $\varphi$ acquires a vacuum expectation value. In view of recent applications of Bose--Einstein condensation, which is a phase transition with an order parameter of the same type, these are particularly interesting results. They may help justifying scenarii which have been recently proposed in the GFT literature, with important applications in cosmology~\cite{gfc_letter, gfc_review, gfc_bounce} and black hole physics~\cite{gfc_bh}.

Non-Abelian models may also be explored in this formalism, and a particularly interesting one is again the $d=3$ theory on ${\rm SU}(2)$ \cite{cor_su2}. Checking its perturbative behaviour requires to push the truncation to order $6$ at least, and even to higher orders if one wants to also account for perturbative $2$-loop contributions. Moreover, the Litim cut-of\/f is not very convenient in this setting, because of the gauge invariant condition. There was no dif\/f\/iculty in the Abelian model mentioned before because the closure constraint translates into a simple momentum constraint $p_1 + \dots + p_6 = 0$, which can be explicitly dealt with. With ${\rm SU}(2)$ the situation is not so simple: the gauge invariance condition encodes complicated recoupling relations among the~${\rm SU}(2)$ harmonic modes, and results in quite challenging expressions. It is therefore better to work in direct space and to rely on a heat-kernel regularization, as was already done in the perturbative Section~\ref{sec:pert}. Such a construction is under way~[Carrozza S., Lahoche V., work {i}n progress]: even if the f\/low equations obtained within this renormalization scheme are not algebraic fractions, they are nonetheless computable and can be integrated out numerically. We should therefore soon be able to directly compute the properties of the f\/low of model~\cite{cor_su2} in a~$\varphi^6$ truncation, and compare them with the features we extrapolated from the formal $\varepsilon$-expansion.

\section{Summary and outlook}

We hope we have managed to convince the reader that GFT renormalization is an active f\/ield of research which has already born interesting fruits. For one thing, the fact that such non-local f\/ield theories can be def\/ined and analyzed by means of renormalization methods may at f\/irst sound like a contradiction in the terms. Locality is indeed a primary concept in relativistic quantum f\/ield theories, and is absolutely key to the formulation of renormalization theory. While at the fundamental level, GFT in a sense goes away with space-time altogether, it is remarkable that tensor invariance may successfully be used as a substitute for locality. On top of providing a (for a long time missing) structure encoding the topology of GFT interactions and Feynman diagrams, it introduces just enough of f\/lexibility to allow for a GFT theory space which is stable under renormalization.

The inclusion of the GFT gauge invariance condition (\ref{gauge_inv}) into renormalizable tensor f\/ield theories was a necessary and by no means obvious step in the direction of quantum gravity applications. We have explained in some detail the remarkable interplay between topology and renormalization which makes it possible (Section~\ref{tracial_2compl}): in perturbative expansion, the most divergent spin foam amplitudes turn out to be supported on simply connected $2$-complexes, which allows to trivialize the bulk holonomies associated to ultraviolet scales, and therefore reabsorb the associated divergences into ef\/fective tensor invariant coupling constants.

This produced a rather large class of renormalizable TGFTs with closure constraint (Tab\-le~\ref{melonic_models}), and consequently a natural test bed for GFT renormalization. Renormalization group stu\-dies have in particular shown that asymptotic freedom is a generic feature of $\varphi^4$ perturbative models, which may therefore be interpreted as ultraviolet complete theories. Moreover, the non-perturbative truncations investigated so far suggest that non-trivial f\/ixed points with properties analogous to that of the Wilson--Fisher f\/ixed point of local scalar f\/ield theories are generic. This opens the way to the study of GFT phases and phase transitions.

We conclude with a non-exhaustive list of open problems which we consider particularly interesting.

{\it Inclusion of non-melonic interactions.} The renormalizable models presented in this review are governed by melonic interactions and melonic radiative corrections. We have however pointed out on several occasions that non-melonic bubbles (such as the necklace bubble of Fig.~\ref{necklace_2point}) may also be included and potentially lead to the def\/inition of yet other perturbative phases. This question deserves to be explored more systematically, as it should in particular help us understand to which extent the polymer phases typically generated by melonic families of graphs -- which have a tree-like structure -- may be escaped \cite{melon_branched}.

{\it Local potential approximation and beyond.} A heuristic derivation of an extended tensor theory space, which includes interactions with arbitrary $G$-invariant dif\/ferential operators, has been proposed in Section~\ref{heuristic_tt}. This suggests to interpret all the renormalizable models studied so far as local potential approximations within this extended tensor theory space, and therefore to explore the properties of more general truncations. Our heuristic argument also shows that in $3d$ and with $G = {\rm SU}(2)$, the generalized tensor theory space contains in principle the colored Boulatov model. This might open the way to a proper quantum gravity interpretation of this three-dimensional theory.

\looseness=-1
{\it Lorentzian signature.} The whole literature on GFT and GFT renormalization is focused on models with compact Lie groups, which may at best result in consistent formulations of Euclidean quantum gravity. The GFT formulation of quantum gravity models with Lorentz signature necessitates to go beyond this framework. While a renormalization scheme taking the additional infrared divergences associated to non-compact linear groups is already availab\-le~\cite{riccardo, frg_r_d}, the physically relevant ${\rm SL}(2, \mathbb{R})$ and ${\rm SL}(2, \mathbb{C})$ are much more challenging. Indeed, the invariant Laplace operators on such groups are not positive, leading to complications associated to the def\/inition of natural propagators. To a large extent, these theories resemble quantum f\/ields on Minkowski space-time, therefore Euclidean multiscale methods are not easily applicable. Alternative techniques, based for instance on Epstein--Glaser renormalization (as it is applicable in any signature) or on a Wick rotation, will be necessary to explore this question in greater detail.

{\it $4d$ quantum gravity constraints.} Eventually one will need to check whether the now well-understood closure constraint can be consistently complemented with spin foam simplicity constraints. If a renormalizable $4d$ model can be def\/ined in this context, we will obtain a consistent perturbative sum over spin foam transition amplitudes, and therefore a tentative def\/inition of the dynamics of LQG. All the renormalization group methods that have been developed for and tested on our simpler toy-models will then be of great practical use, in particular to determine whether a sector of the quantum dynamics reproduces general relativity in a suitably def\/ined classical limit.

\subsection*{Acknowledgements}

The author acknowledges support from the ANR JCJC CombPhysMat2Tens grant.

\pdfbookmark[1]{References}{ref}
\LastPageEnding


\begin{thebibliography}{99}
\footnotesize\itemsep=0pt

\bibitem{Alexandrov:2011ab}
Alexandrov S., Geiller M., Noui K., Spin foams and canonical quantization,
 \href{http://dx.doi.org/10.3842/SIGMA.2012.055}{\textit{SIGMA}} \textbf{8} (2012), 055, 79~pages, \href{http://arxiv.org/abs/1112.1961}{arXiv:1112.1961}.

\bibitem{ashtekar1995differential}
Ashtekar A., Lewandowski J., Dif\/ferential geometry on the space of connections
 via graphs and projective limits, \href{http://dx.doi.org/10.1016/0393-0440(95)00028-G}{\textit{J.~Geom. Phys.}} \textbf{17} (1995),
 191--230, \href{http://arxiv.org/abs/hep-th/9412073}{hep-th/9412073}.

\bibitem{review_frg}
Bagnuls C., Bervillier C., Exact renormalization group equations: an
 introductory review, \href{http://dx.doi.org/10.1016/S0370-1573(00)00137-X}{\textit{Phys. Rep.}} \textbf{348} (2001), 91--157,
 \href{http://arxiv.org/abs/hep-th/0002034}{hep-th/0002034}.

\bibitem{bahr2014}
Bahr B., On background-independent renormalization of spin foam models,
 \href{http://arxiv.org/abs/1407.7746}{arXiv:1407.7746}.

\bibitem{melonic_phase}
Baratin A., Carrozza S., Oriti D., Ryan J., Smerlak M., Melonic phase
 transition in group f\/ield theory, \href{http://dx.doi.org/10.1007/s11005-014-0699-9}{\textit{Lett. Math. Phys.}} \textbf{104}
 (2014), 1003--1017, \href{http://arxiv.org/abs/1307.5026}{arXiv:1307.5026}.

\bibitem{diffeos}
Baratin A., Girelli F., Oriti D., Dif\/feomorphisms in group f\/ield theories,
 \href{http://dx.doi.org/10.1103/PhysRevD.83.104051}{\textit{Phys. Rev.~D}} \textbf{83} (2011), 104051, 22~pages,
 \href{http://arxiv.org/abs/1101.0590}{arXiv:1101.0590}.

\bibitem{bo_bc}
Baratin A., Oriti D., Quantum simplicial geometry in the group f\/ield theory
 formalism: reconsidering the Barrett--Crane model, \href{http://dx.doi.org/10.1088/1367-2630/13/12/125011}{\textit{New~J. Phys.}}
 \textbf{13} (2011), 125011, 28~pages, \href{http://arxiv.org/abs/1108.1178}{arXiv:1108.1178}.

\bibitem{bo}
Baratin A., Oriti D., Group f\/ield theory and simplicial gravity path integrals:
 a model for Holst--Plebanski gravity, \href{http://dx.doi.org/10.1103/PhysRevD.85.044003}{\textit{Phys. Rev.~D}} \textbf{85}
 (2012), 044003, 15~pages, \href{http://arxiv.org/abs/1111.5842}{arXiv:1111.5842}.

\bibitem{ad_ten}
Baratin A., Oriti D., Ten questions on group f\/ield theory (and their tentative
 answers), \href{http://dx.doi.org/10.1088/1742-6596/360/1/012002}{\textit{J.~Phys. Conf. Ser.}} \textbf{360} (2012), 012002, 10~pages,
 \href{http://arxiv.org/abs/1112.3270}{arXiv:1112.3270}.

\bibitem{josephaf}
Ben~Geloun J., Two- and four-loop {$\beta$}-functions of rank-4 renormalizable
 tensor f\/ield theories, \href{http://dx.doi.org/10.1088/0264-9381/29/23/235011}{\textit{Classical Quantum Gravity}} \textbf{29} (2012),
 235011, 40~pages, \href{http://arxiv.org/abs/1205.5513}{arXiv:1205.5513}.

\bibitem{joseph_d2}
Ben~Geloun J., Renormalizable models in rank {$d\geq 2$} tensorial group f\/ield
 theory, \href{http://dx.doi.org/10.1007/s00220-014-2142-6}{\textit{Comm. Math. Phys.}} \textbf{332} (2014), 117--188,
 \href{http://arxiv.org/abs/1306.1201}{arXiv:1306.1201}.

\bibitem{Geloun:2015lta}
Ben~Geloun J., A power counting theorem for a $p^{2a}\phi^4$ tensorial group
 f\/ield theory, \href{http://arxiv.org/abs/1507.00590}{arXiv:1507.00590}.

\bibitem{Geloun:2010vj}
Ben~Geloun J., Gurau R., Rivasseau V., EPRL/FK group f\/ield theory,
 \href{http://dx.doi.org/10.1209/0295-5075/92/60008}{\textit{Europhys. Lett.}} \textbf{92} (2010), 60008, 6~pages,
 \href{http://arxiv.org/abs/1008.0354}{arXiv:1008.0354}.

\bibitem{lin_gft}
Ben~Geloun J., Krajewski T., Magnen J., Rivasseau V., Linearized group f\/ield
 theory and power-counting theorems, \href{http://dx.doi.org/10.1088/0264-9381/27/15/155012}{\textit{Classical Quantum Gravity}}
 \textbf{27} (2010), 155012, 14~pages, \href{http://arxiv.org/abs/1002.3592}{arXiv:1002.3592}.

\bibitem{joseph_etera}
Ben~Geloun J., Livine E.R., Some classes of renormalizable tensor models,
 \href{http://dx.doi.org/10.1063/1.4818797}{\textit{J.~Math. Phys.}} \textbf{54} (2013), 082303, 25~pages,
 \href{http://arxiv.org/abs/1207.0416}{arXiv:1207.0416}.

\bibitem{riccardo}
Ben~Geloun J., Martini R., Oriti D., Functional renormalization group analysis
 of a tensorial group f\/ield theory on $\mathbb{R}^3$, \href{http://dx.doi.org/10.1209/0295-5075/112/31001}{\textit{Europhys. Lett.}}
 \textbf{112} (2015), 31001, 6~pages, \href{http://arxiv.org/abs/1508.01855}{arXiv:1508.01855}.

\bibitem{frg_r_d}
Ben~Geloun J., Martini R., Oriti D., Functional renormalisation group analysis
 of tensorial group f\/ield theories on $\mathbb{R}^d$, \href{http://dx.doi.org/10.1103/PhysRevD.94.024017}{\textit{Phys. Rev.~D}}
 \textbf{94} (2016), 024017, 45~pages, \href{http://arxiv.org/abs/1601.08211}{arXiv:1601.08211}.

\bibitem{tensor_4d}
Ben~Geloun J., Rivasseau V., A renormalizable 4-dimensional tensor f\/ield
 theory, \href{http://dx.doi.org/10.1007/s00220-012-1549-1}{\textit{Comm. Math. Phys.}} \textbf{318} (2013), 69--109,
 \href{http://arxiv.org/abs/1111.4997}{arXiv:1111.4997}.

\bibitem{addendum}
Ben~Geloun J., Rivasseau V., Addendum to: {A} renormalizable 4-dimensional
 tensor f\/ield theory, \href{http://dx.doi.org/10.1007/s00220-013-1703-4}{\textit{Comm. Math. Phys.}} \textbf{322} (2013),
 957--965, \href{http://arxiv.org/abs/1209.4606}{arXiv:1209.4606}.

\bibitem{josephsamary}
Ben~Geloun J., Samary D.O., 3{D} tensor f\/ield theory: renormalization and
 one-loop {$\beta$}-functions, \href{http://dx.doi.org/10.1007/s00023-012-0225-5}{\textit{Ann. Henri Poincar\'e}} \textbf{14}
 (2013), 1599--1642, \href{http://arxiv.org/abs/1201.0176}{arXiv:1201.0176}.

\bibitem{joseph_reiko}
Ben~Geloun J., Toriumi R., Parametric representation of rank {$d$} tensorial
 group f\/ield theory: {A}belian models with kinetic term {$\sum_s\vert p_s\vert
 +\mu$}, \href{http://dx.doi.org/10.1063/1.4929771}{\textit{J.~Math. Phys.}} \textbf{56} (2015), 093503, 53~pages,
 \href{http://arxiv.org/abs/1409.0398}{arXiv:1409.0398}.

\bibitem{dario_vincentL}
Benedetti D., Lahoche V., Functional renormalization group approach for
 tensorial group f\/ield theo\-ry: a~rank-6 model with closure constraint,
 \href{http://dx.doi.org/10.1088/0264-9381/33/9/095003}{\textit{Classical Quantum Gravity}} \textbf{33} (2016), 095003, 35~pages,
 \href{http://arxiv.org/abs/1508.06384}{arXiv:1508.06384}.

\bibitem{valentin_enhance}
Bonzom V., Delepouve T., Rivasseau V., Enhancing non-melonic triangulations: a
 tensor model mixing melonic and planar maps, \href{http://dx.doi.org/10.1016/j.nuclphysb.2015.04.004}{\textit{Nuclear Phys.~B}}
 \textbf{895} (2015), 161--191, \href{http://arxiv.org/abs/1502.01365}{arXiv:1502.01365}.

\bibitem{critical}
Bonzom V., Gurau R., Riello A., Rivasseau V., Critical behavior of colored
 tensor models in the large~{$N$} limit, \href{http://dx.doi.org/10.1016/j.nuclphysb.2011.07.022}{\textit{Nuclear Phys.~B}} \textbf{853}
 (2011), 174--195, \href{http://arxiv.org/abs/1105.3122}{arXiv:1105.3122}.

\bibitem{uncoloring}
Bonzom V., Gurau R., Rivasseau V., Random tensor models in the large $N$ limit:
 uncoloring the colored tensor models, \href{http://dx.doi.org/10.1103/PhysRevD.85.084037}{\textit{Phys. Rev.~D}} \textbf{85}
 (2012), 084037, 12~pages, \href{http://arxiv.org/abs/1202.3637}{arXiv:1202.3637}.

\bibitem{double_scaling}
Bonzom V., Gurau R., Ryan J.P., Tanasa A., The double scaling limit of random
 tensor models, \href{http://dx.doi.org/10.1007/JHEP09(2014)051}{\textit{J.~High Energy Phys.}} \textbf{2014} (2014), no.~9,
 051, 49~pages, \href{http://arxiv.org/abs/1404.7517}{arXiv:1404.7517}.

\bibitem{vm1}
Bonzom V., Smerlak M., Bubble divergences from cellular cohomology,
 \href{http://dx.doi.org/10.1007/s11005-010-0414-4}{\textit{Lett. Math. Phys.}} \textbf{93} (2010), 295--305, \href{http://arxiv.org/abs/1004.5196}{arXiv:1004.5196}.

\bibitem{vm3}
Bonzom V., Smerlak M., Bubble divergences: sorting out topology from cell
 structure, \href{http://dx.doi.org/10.1007/s00023-011-0127-y}{\textit{Ann. Henri Poincar\'e}} \textbf{13} (2012), 185--208,
 \href{http://arxiv.org/abs/1103.3961}{arXiv:1103.3961}.

\bibitem{boulatov}
Boulatov D.V., A model of three-dimensional lattice gravity, \href{http://dx.doi.org/10.1142/S0217732392001324}{\textit{Modern
 Phys. Lett.~A}} \textbf{7} (1992), 1629--1646, \href{http://arxiv.org/abs/hep-th/9202074}{hep-th/9202074}.

\bibitem{thesis}
Carrozza S., Tensorial methods and renormalization in group f\/ield theories,
 \href{http://dx.doi.org/10.1007/978-3-319-05867-2}{\textit{Springer Theses}}, Springer, Cham, 2014, \href{http://arxiv.org/abs/1310.3736}{arXiv:1310.3736}.

\bibitem{discrete_rg}
Carrozza S., Discrete renormalization group for {${\rm SU}(2)$} tensorial group
 f\/ield theory, \href{http://dx.doi.org/10.4171/AIHPD/15}{\textit{Ann. Inst. Henri Poincar\'e~D}} \textbf{2} (2015),
 49--112, \href{http://arxiv.org/abs/1407.4615}{arXiv:1407.4615}.

\bibitem{4-eps}
Carrozza S., Group f\/ield theory in dimension {$4-\varepsilon$}, \href{http://dx.doi.org/10.1103/PhysRevD.91.065023}{\textit{Phys.
 Rev.~D}} \textbf{91} (2015), 065023, 10~pages, \href{http://arxiv.org/abs/1411.5385}{arXiv:1411.5385}.

\bibitem{bubbles}
Carrozza S., Oriti D., Bounding bubbles: the vertex representation of $3d$ group
 f\/ield theory and the suppression of pseudomanifolds, \href{http://dx.doi.org/10.1103/PhysRevD.85.044004}{\textit{Phys. Rev.~D}}
 \textbf{85} (2012), 044004, 22~pages, \href{http://arxiv.org/abs/1104.5158}{arXiv:1104.5158}.

\bibitem{ooguri_edge}
Carrozza S., Oriti D., Bubbles and jackets: new scaling bounds in topological
 group f\/ield theories, \href{http://dx.doi.org/10.1007/JHEP06(2012)092}{\textit{J.~High Energy Phys.}} \textbf{2012} (2012),
 no.~6, 092, 42~pages, \href{http://arxiv.org/abs/1203.5082}{arXiv:1203.5082}.

\bibitem{cor_su2}
Carrozza S., Oriti D., Rivasseau V., Renormalization of a {${\rm SU}(2)$}
 tensorial group f\/ield theory in three dimensions, \href{http://dx.doi.org/10.1007/s00220-014-1928-x}{\textit{Comm. Math. Phys.}}
 \textbf{330} (2014), 581--637, \href{http://arxiv.org/abs/1303.6772}{arXiv:1303.6772}.

\bibitem{cor_u1}
Carrozza S., Oriti D., Rivasseau V., Renormalization of tensorial group f\/ield
 theories: {A}belian {${\rm U}(1)$} models in four dimensions, \href{http://dx.doi.org/10.1007/s00220-014-1954-8}{\textit{Comm.
 Math. Phys.}} \textbf{327} (2014), 603--641, \href{http://arxiv.org/abs/1207.6734}{arXiv:1207.6734}.

\bibitem{dPFKR}
De~Pietri R., Freidel L., Krasnov K., Rovelli C., Barrett--{C}rane model from a
 {B}oulatov--{O}oguri f\/ield theory over a homogeneous space, \href{http://dx.doi.org/10.1016/S0550-3213(00)00005-5}{\textit{Nuclear
 Phys.~B}} \textbf{574} (2000), 785--806, \href{http://arxiv.org/abs/hep-th/9907154}{hep-th/9907154}.

\bibitem{bianca_continuum_2014}
Dittrich B., The continuum limit of loop quantum gravity~-- a~framework for
 solving the theory, \href{http://arxiv.org/abs/1409.1450}{arXiv:1409.1450}.

\bibitem{bianca_marc_new}
Dittrich B., Geiller M., A new vacuum for loop quantum gravity,
 \href{http://dx.doi.org/10.1088/0264-9381/32/11/112001}{\textit{Classical Quantum Gravity}} \textbf{32} (2015), 112001, 13~pages,
 \href{http://arxiv.org/abs/1401.6441}{arXiv:1401.6441}.

\bibitem{dl}
Dupuis M., Livine E.R., Holomorphic simplicity constraints for 4{D} spinfoam
 models, \href{http://dx.doi.org/10.1088/0264-9381/28/21/215022}{\textit{Classical Quantum Gravity}} \textbf{28} (2011), 215022,
 32~pages, \href{http://arxiv.org/abs/1104.3683}{arXiv:1104.3683}.

\bibitem{eprl}
Engle J., Livine E., Pereira R., Rovelli C., L{QG} vertex with f\/inite {I}mmirzi
 parameter, \href{http://dx.doi.org/10.1016/j.nuclphysb.2008.02.018}{\textit{Nuclear Phys.~B}} \textbf{799} (2008), 136--149,
 \href{http://arxiv.org/abs/0711.0146}{arXiv:0711.0146}.

\bibitem{ferri1986}
Ferri M., Gagliardi C., Grasselli L., A graph-theoretical representation of
 {PL}-manifolds~-- a~survey on crystallizations, \href{http://dx.doi.org/10.1007/BF02188181}{\textit{Aequationes Math.}}
 \textbf{31} (1986), 121--141.

\bibitem{freidel_gft}
Freidel L., Group f\/ield theory: an overview, \href{http://dx.doi.org/10.1007/s10773-005-8894-1}{\textit{Internat.~J. Theoret.
 Phys.}} \textbf{44} (2005), 1769--1783, \mbox{\href{http://arxiv.org/abs/hep-th/0505016}{hep-th/0505016}}.

\bibitem{lrd_ren}
Freidel L., Gurau R., Oriti D., Group f\/ield theory renormalization in the 3D
 case: power counting of divergences, \href{http://dx.doi.org/10.1103/PhysRevD.80.044007}{\textit{Phys. Rev.~D}} \textbf{80}
 (2009), 044007, 20~pages, \href{http://arxiv.org/abs/0905.3772}{arXiv:0905.3772}.

\bibitem{fk}
Freidel L., Krasnov K., A new spin foam model for 4{D} gravity,
 \href{http://dx.doi.org/10.1088/0264-9381/25/12/125018}{\textit{Classical Quantum Gravity}} \textbf{25} (2008), 125018, 36~pages,
 \href{http://arxiv.org/abs/0708.1595}{arXiv:0708.1595}.

\bibitem{freidel_louapre_PRI}
Freidel L., Louapre D., Ponzano--{R}egge model revisited. {I}.~{G}auge f\/ixing,
 observables and interacting spinning particles, \href{http://dx.doi.org/10.1088/0264-9381/21/24/002}{\textit{Classical Quantum
 Gravity}} \textbf{21} (2004), 5685--5726, \href{http://arxiv.org/abs/hep-th/0401076}{hep-th/0401076}.

\bibitem{gfc_letter}
Gielen S., Oriti D., Sindoni L., Cosmology from group f\/ield theory formalism
 for quantum gravity, \href{http://dx.doi.org/10.1103/PhysRevLett.111.031301}{\textit{Phys. Rev. Lett.}} \textbf{111} (2013), 031301,
 4~pages, \href{http://arxiv.org/abs/1303.3576}{arXiv:1303.3576}.

\bibitem{gfc_review}
Gielen S., Sindoni L., Quantum cosmology from group f\/ield theory condensates:
 a~review, \href{http://arxiv.org/abs/1602.08104}{arXiv:1602.08104}.

\bibitem{razvan_colors}
Gurau R., Colored group f\/ield theory, \href{http://dx.doi.org/10.1007/s00220-011-1226-9}{\textit{Comm. Math. Phys.}} \textbf{304}
 (2011), 69--93, \href{http://arxiv.org/abs/0907.2582}{arXiv:0907.2582}.

\bibitem{virasoro}
Gurau R., A generalization of the {V}irasoro algebra to arbitrary dimensions,
 \href{http://dx.doi.org/10.1016/j.nuclphysb.2011.07.009}{\textit{Nuclear Phys.~B}} \textbf{852} (2011), 592--614, \href{http://arxiv.org/abs/1105.6072}{arXiv:1105.6072}.

\bibitem{razvanN}
Gurau R., The {$1/N$} expansion of colored tensor models, \href{http://dx.doi.org/10.1007/s00023-011-0101-8}{\textit{Ann. Henri
 Poincar\'e}} \textbf{12} (2011), 829--847, \href{http://arxiv.org/abs/1011.2726}{arXiv:1011.2726}.

\bibitem{razvan_complete}
Gurau R., The complete {$1/N$} expansion of colored tensor models in arbitrary
 dimension, \href{http://dx.doi.org/10.1007/s00023-011-0118-z}{\textit{Ann. Henri Poincar\'e}} \textbf{13} (2012), 399--423,
 \href{http://arxiv.org/abs/1102.5759}{arXiv:1102.5759}.

\bibitem{universality}
Gurau R., Universality for random tensors, \href{http://dx.doi.org/10.1214/13-AIHP567}{\textit{Ann. Inst. Henri Poincar\'e
 Probab. Stat.}} \textbf{50} (2014), 1474--1525, \href{http://arxiv.org/abs/1111.0519}{arXiv:1111.0519}.

\bibitem{RazvanVincentN}
Gurau R., Rivasseau V., The $1/N$ expansion of colored tensor models in
 arbitrary dimension, \href{http://dx.doi.org/10.1209/0295-5075/95/50004}{\textit{Europhys. Lett.}} \textbf{95} (2011), 50004,
 5~pages, \href{http://arxiv.org/abs/1101.4182}{arXiv:1101.4182}.

\bibitem{razvan_jimmy_rev}
Gurau R., Ryan J.P., Colored tensor models~-- a~review, \href{http://dx.doi.org/10.3842/SIGMA.2012.020}{\textit{SIGMA}}
 \textbf{8} (2012), 020, 78~pages, \href{http://arxiv.org/abs/1109.4812}{arXiv:1109.4812}.

\bibitem{melon_branched}
Gurau R., Ryan J.P., Melons are branched polymers, \href{http://dx.doi.org/10.1007/s00023-013-0291-3}{\textit{Ann. Henri
 Poincar\'e}} \textbf{15} (2014), 2085--2131, \href{http://arxiv.org/abs/1302.4386}{arXiv:1302.4386}.

\bibitem{Krajewski_rev}
Krajewski T., Group f\/ield theories, \textit{PoS Proc. Sci.} (2011),
 PoS(QGQGS2011), 005, 58~pages, \href{http://arxiv.org/abs/1210.6257}{arXiv:1210.6257}.

\bibitem{Krajewski:2010yq}
Krajewski T., Magnen J., Rivasseau V., Tanasa A., Vitale P., Quantum
 corrections in the group f\/ield theory formulation of the
 Engle--Pereira--Rovelli--Livine and Freidel--Krasnov models, \href{http://dx.doi.org/10.1103/PhysRevD.82.124069}{\textit{Phys.
 Rev.~D}} \textbf{82} (2010), 124069, 20~pages, \href{http://arxiv.org/abs/1007.3150}{arXiv:1007.3150}.

\bibitem{polch_gft}
Krajewski T., Toriumi R., Polchinski's equation for group f\/ield theory,
 \href{http://dx.doi.org/10.1002/prop.201400043}{\textit{Fortschr. Phys.}} \textbf{62} (2014), 855--862.

\bibitem{thomas_reiko}
Krajewski T., Toriumi R., Polchinski's exact renormalisation group for
 tensorial theories: {G}aussian universality and power counting,
 \href{http://arxiv.org/abs/1511.09084}{arXiv:1511.09084}.

\bibitem{thomas_reiko_sigma}
Krajewski T., Toriumi R., Exact renormalisation group equations and loop
 equations for tensor models, \href{http://dx.doi.org/10.3842/SIGMA.2016.068}{\textit{SIGMA}} \textbf{12} (2016), 068,
 36~pages, \href{http://arxiv.org/abs/1603.00172}{arXiv:1603.00172}.

\bibitem{vincentL_constructive}
Lahoche V., Constructive tensorial group f\/ield theory~{I}: the $U(1)-T^4_3$
 model, \href{http://arxiv.org/abs/1510.05050}{arXiv:1510.05050}.

\bibitem{vincentL_constructive2}
Lahoche V., Constructive tensorial group f\/ield theory~{II}: the $U(1)-T^4_4$
 model, \href{http://arxiv.org/abs/1510.05051}{arXiv:1510.05051}.

\bibitem{vincent2_daniele}
Lahoche V., Oriti D., Rivasseau V., Renormalization of an {A}belian tensor
 group f\/ield theory: solution at leading order, \href{http://dx.doi.org/10.1007/JHEP04(2015)095}{\textit{J.~High Energy Phys.}}
 \textbf{2015} (2015), no.~4, 095, 41~pages, \href{http://arxiv.org/abs/1501.02086}{arXiv:1501.02086}.

\bibitem{litim}
Litim D.F., Optimization of the exact renormalization group, \href{http://dx.doi.org/10.1016/S0370-2693(00)00748-6}{\textit{Phys.
 Lett.~B}} \textbf{486} (2000), 92--99, \mbox{\href{http://arxiv.org/abs/hep-th/0005245}{hep-th/0005245}}.

\bibitem{matteo_scaling}
Magnen J., Noui K., Rivasseau V., Smerlak M., Scaling behavior of
 three-dimensional group f\/ield theory, \href{http://dx.doi.org/10.1088/0264-9381/26/18/185012}{\textit{Classical Quantum Gravity}}
 \textbf{26} (2009), 185012, 25~pages, \href{http://arxiv.org/abs/0906.5477}{arXiv:0906.5477}.

\bibitem{morris}
Morris T.R., The exact renormalization group and approximate solutions,
 \href{http://dx.doi.org/10.1142/S0217751X94000972}{\textit{Internat.~J. Modern Phys.~A}} \textbf{9} (1994), 2411--2450,
 \href{http://arxiv.org/abs/hep-ph/9308265}{hep-ph/9308265}.

\bibitem{ooguri}
Ooguri H., Topological lattice models in four dimensions, \href{http://dx.doi.org/10.1142/S0217732392004171}{\textit{Modern Phys.
 Lett.~A}} \textbf{7} (1992), 2799--2810, \mbox{\href{http://arxiv.org/abs/hep-th/9205090}{hep-th/9205090}}.

\bibitem{daniele_hydro}
Oriti D., Group f\/ield theory as the microscopic description of the quantum
 spacetime f\/luid: a new perspective on the continuum in quantum gravity,
 \textit{PoS Proc. Sci.} (2007), PoS(QG--Ph), 030, 38~pages,
 \href{http://arxiv.org/abs/0710.3276}{arXiv:0710.3276}.

\bibitem{daniele_rev2006}
Oriti D., The group f\/ield theory approach to quantum gravity, in Approaches to
 Quantum Gravity~-- toward a New Understanding of Space, Time, and Matter,
 Cambridge University Press, Cambridge, 2009, 310--331,
 \href{http://arxiv.org/abs/gr-qc/0607032}{gr-qc/0607032}.

\bibitem{daniele_rev2011}
Oriti D., The microscopic dynamics of quantum space as a group f\/ield theory, in
 Foundations of Space and Time, Cambridge University Press, Cambridge, 2012,
 257--320, \href{http://arxiv.org/abs/1110.5606}{arXiv:1110.5606}.

\bibitem{gfc_bh}
Oriti D., Pranzetti D., Sindoni L., Horizon entropy from quantum gravity
 condensates, \href{http://dx.doi.org/10.1103/PhysRevLett.116.211301}{\textit{Phys. Rev. Lett.}} \textbf{116} (2016), 211301, 6~pages,
 \href{http://arxiv.org/abs/1510.06991}{arXiv:1510.06991}.

\bibitem{gfc_bounce}
Oriti D., Sindoni L., Wilson-Ewing E., Bouncing cosmologies from quantum
 gravity condensates, \href{http://arxiv.org/abs/1602.08271}{arXiv:1602.08271}.

\bibitem{perez_review2012}
Perez A., The spin foam approach to quantum gravity, \href{http://dx.doi.org/10.12942/lrr-2013-3}{\textit{Living Rev.
 Relativ.}} \textbf{16} (2013), 3, 128~pages, \href{http://arxiv.org/abs/1205.2019}{arXiv:1205.2019}.

\bibitem{GFT_rovelli_reisenberg}
Reisenberger M.P., Rovelli C., Spacetime as a {F}eynman diagram: the connection
 formulation, \href{http://dx.doi.org/10.1088/0264-9381/18/1/308}{\textit{Classical Quantum Gravity}} \textbf{18} (2001), 121--140,
 \href{http://arxiv.org/abs/gr-qc/0002095}{gr-qc/0002095}.

\bibitem{aldo_eprl}
Riello A., Self-energy of the Lorentzian Engle--Pereira--Rovelli--Livine and
 Freidel--Krasnov model of quantum gravity, \href{http://dx.doi.org/10.1103/PhysRevD.88.024011}{\textit{Phys. Rev.~D}} \textbf{88}
 (2013), 024011, 30~pages, \href{http://arxiv.org/abs/1302.1781}{arXiv:1302.1781}.

\bibitem{vincent_book}
Rivasseau V., From perturbative to constructive renormalization, \href{http://dx.doi.org/10.1515/9781400862085}{\textit{Princeton
 Series in Physics}}, Princeton University Press, Princeton, NJ, 1991.

\bibitem{tt1}
Rivasseau V., Quantum gravity and renormalization: the tensor track,
 \href{http://dx.doi.org/10.1063/1.4715396}{\textit{AIP Conf. Proc.}} \textbf{1444} (2012), 18--29, \href{http://arxiv.org/abs/1112.5104}{arXiv:1112.5104}.

\bibitem{vincent_af}
Rivasseau V., Why are tensor f\/ield theories asymptotically free?,
 \href{http://dx.doi.org/10.1209/0295-5075/111/60011}{\textit{Europhys. Lett.}} \textbf{111} (2015), 60011, 6~pages,
 \href{http://arxiv.org/abs/1507.04190}{arXiv:1507.04190}.

\bibitem{rovelli_book}
Rovelli C., Quantum gravity, \href{http://dx.doi.org/10.1017/CBO9780511755804}{\textit{Cambridge Monographs on Mathematical Physics}},
 Cambridge University Press, Cambridge, 2004.

\bibitem{samary_beta}
Samary D.O., Beta functions of ${\rm U}(1)^d$ gauge invariant just
 renormalizable tensor models, \href{http://dx.doi.org/10.1103/PhysRevD.88.105003}{\textit{Phys. Rev.~D}} \textbf{88} (2013),
 105003, 15~pages, \href{http://arxiv.org/abs/1303.7256}{arXiv:1303.7256}.

\bibitem{samary_2point}
Samary D.O., Closed equations of the two-point functions for tensorial group
 f\/ield theory, \href{http://dx.doi.org/10.1088/0264-9381/31/18/185005}{\textit{Classical Quantum Gravity}} \textbf{31} (2014), 185005,
 29~pages, \href{http://arxiv.org/abs/1401.2096}{arXiv:1401.2096}.

\bibitem{dine_fabien_2015}
Samary D.O., P{\'e}rez-S{\'a}nchez C.I., Vignes-Tourneret F., Wulkenhaar R.,
 Correlation functions of a just renormalizable tensorial group f\/ield theory:
 the melonic approximation, \href{http://dx.doi.org/10.1088/0264-9381/32/17/175012}{\textit{Classical Quantum Gravity}} \textbf{32}
 (2015), 175012, 18~pages, \href{http://arxiv.org/abs/1411.7213}{arXiv:1411.7213}.

\bibitem{samary_vignes}
Samary D.O., Vignes-Tourneret F., Just renormalizable {TGFT}'s on {${\rm
 U}(1)^d$} with gauge invariance, \href{http://dx.doi.org/10.1007/s00220-014-1930-3}{\textit{Comm. Math. Phys.}} \textbf{329}
 (2014), 545--578, \href{http://arxiv.org/abs/1211.2618}{arXiv:1211.2618}.

\bibitem{thiemann_book}
Thiemann T., Modern canonical quantum general relativity, \href{http://dx.doi.org/10.1017/CBO9780511755682}{\textit{Cambridge Monographs
 on Mathematical Physics}}, Cambridge University Press, Cambridge, 2007,
 \href{http://arxiv.org/abs/gr-qc/0110034}{gr-qc/0110034}.

\bibitem{wetterich_eq}
Wetterich C., Exact evolution equation for the ef\/fective potential,
 \href{http://dx.doi.org/10.1016/0370-2693(93)90726-X}{\textit{Phys. Lett.~B}} \textbf{301} (1993), 90--94.

\bibitem{wilson1974renormalization}
Wilson K.G., Kogut J., The renormalization group and the $\epsilon$ expansion,
 \href{http://dx.doi.org/10.1016/0370-1573(74)90023-4}{\textit{Phys. Rep.}} \textbf{12} (1974), 75--199.

\end{thebibliography}
\end{document}